\documentclass[a4paper,11pt,twoside]{amsart}

\usepackage[T1]{fontenc}
\usepackage{amsfonts,amsmath,amssymb,amsxtra,amscd}
\usepackage{mathrsfs,ae,dsfont}
\usepackage{graphicx,color}
\usepackage[multiple]{footmisc}
\usepackage{layout}
\usepackage[nice]{nicefrac}

\def\plus#1#2{\vrule height#1pt width0pt depth#2pt}
\def\@{\hskip.8pt}
\def\?{\hskip.3pt}
\def\And{,\@\ldots\hskip-.4pt,}

\newcommand{\rarw}[1]{\overset{#1}{\longrightarrow}}

\renewcommand{\leq}{\leqslant}
\renewcommand{\geq}{\geqslant}
\newcommand{\h}[1]{\hat{#1}}
\newcommand{\ovl}[1]{\overline{#1}}
\renewcommand{\j}[1]{j_1\/(#1)}
\newcommand{\narc}[1]{\text{\@\tiny$(#1)$}}
\newcommand{\interior}{\,\hbox{\vrule depth0pt height.6pt width4pt\vrule depth0pt height8pt}\;\,}
\renewcommand{\vert}[2]{\@\vrule height#1pt width.3pt depth#2pt\@}

\DeclareMathAlphabet{\mathpzc}{OT1}{pzc}{m}{it}

\def\d#1/d#2{\frac{d\/#1}{d\/#2}}
\def\D#1/D#2{\frac{D\/#1}{D\/#2}}
\def\de#1/de#2{\frac{\partial\/#1}{\partial\/#2}}
\def\sd#1/de#2/de#3{\ifx#2 \frac{\plus02\partial^{\@\@2}#1}{\plus90\partial\@#3^{\@2}}%
\else\frac{\plus02\partial^{\@\@2}#1}{\partial\?#2\partial\?#3}\fi}

\def\A{\mathcal A\/}
\def\Ag{A\?(\h\g)}
\def\B{\mathfrak B\/}
\def\C{\mathcal C\/}
\def\DE{\tilde\partial\?}
\def\F{\mathscr{F\/}}
\def\H{\mathcal{H}}
\def\Ham{\mathscr{H}}
\def\Hg{\H\@(\h\g)}
\def\I{\mathcal{I\/}}
\def\K{\mathcal{K}}
\def\Lagr{\mathscr{L}}
\def\O{\mathcal O\/}
\def\P_#1{\mathcal P_{#1}\/}
\def\R{\mathds{R}}
\def\Th{\Theta}
\def\V{\mathcal{V}_{n+1}}
\def\Vg{V\/(\g)}

\def\a{\alpha}
\def\b{\beta}
\def\eps{\varepsilon}
\def\g{\gamma}
\def\gt{{\tilde{\g}}}
\def\k{\kappa}
\def\l{\lambda}
\def\na{\narc{a}}
\def\nb{\narc{b}}
\def\ni{\narc{i}}

\def\s{\sigma}
\def\th{\vartheta}
\def\thL{\th_{\hskip-1pt \Lagr}}
\def\vphi{\varphi}
\def\vsigma{\varsigma}
\def\w{\omega}
\def\wg#1{\@\delta\?q^{#1}{}_{|\g}}
\def\wt{\tilde\w}
\def\z{\zeta}

\def\Ker{\operatorname{ker\@}}

\def\rank{{\rm rank}\,}
\def\const{{\rm const.}}

\newtheorem{remark}{Remark}[section]
\newtheorem{proposition}{Proposition}[section]
\newtheorem{theorem}{Theorem}[section]
\newtheorem{lemma}{Lemma}[section]
\newtheorem{corollary}{Corollary}[section]
\newtheorem{definition}{Definition}[section]

\catcode`\@=11 \@addtoreset{equation}{section}
\def\theequation{\thesection.\arabic{equation}}
\catcode`\@=12

\catcode`\@=11
\renewenvironment{subequations}{%
  \refstepcounter{equation}%
  \protected@edef\theparentequation{\theequation}%
  \setcounter{parentequation}{\value{equation}}%
  \setcounter{equation}{0}%
  \def\theequation{\theparentequation\hspace{1pt}\alph{equation}}%
  \ignorespaces
}{%
  \setcounter{equation}{\value{parentequation}}%
  \ignorespacesafterend
} \catcode`\@=12

\catcode`\@=11
\newcounter{Section}
\refstepcounter{Section} 
\catcode`\@=12
\newtheorem{thm}{Theorem}[Section]

\begin{document}


\title[Constrained variational calculus\@: the second variation\ (\@part\ I\@)]{Constrained variational calculus\@:\\[1pt] the second variation (\@part\ I\@)}

\author[E. Massa et al.]{E. Massa\?,\, D. Bruno\?,\, G. Luria\?,\, E. Pagani} 
\address{Dipartimento di Matematica - Universit\`a di Genova \\
        Via Dodecaneso, 35 - 16146 Genova (Italia) }
\email{massa@dima.unige.it, luria@dima.unige.it}
\address{Department of Advanced Robotics - Istituto Italiano di Tecnologia \\
Via Morego, 30 - 16163 Genova (Italia)}
\email{danilo.bruno\/@\/iit.it}
\address{Dipartimento di Matematica - Universit\`a di Trento \\
       Via Sommarive, 14 - 38050 Povo di Trento (Italia) }
\email{pagani@science.unitn.it}

\begin{abstract}
Within the geometrical framework developed in \cite{mbp1}\?, the problem of minimality for constrained calculus of variations is analysed among the class of
differentiable curves. A fully covariant representation of the second variation of the action functional, based on a suitable gauge transformation of the
Lagrangian, is explicitly worked out. Both necessary and sufficient conditions for minimality are proved, and are then reinterpreted in terms of Jacobi fields.\\


\vspace{2pt}\noindent
\end{abstract}


\maketitle

\vspace{5pt}
\section*{Introduction}\label{S0}
The present paper deals with a geometric approach to constrained calculus of variations and it is aimed at  establishing under which conditions a curve $\g\@$
provides a local minimum for a given action functional $\I\@[\g]\@$.

\vspace{1pt}
A preliminary step in this direction has been taken in \cite{mbp1}, where the first variation of $\I\@[\g]\@$ has  been analysed. This resulted into a set of
conditions characterizing extremal curves among the class of piecewise differentiable ones, i.e.~among the totality  of continuous curves having a finite
number of discontinuities in their first derivative.

In the present work, we shall concentrate on the sub--class of \emph{differentiable\/} curves. All issues arising  from the possible presence of corners are
postponed to a forthcoming paper. In this way, the problem is broken up into two consecutive steps: we now first  seek for the minimality conditions for a
single differentiable arc and then, in the next paper, we shall complete them into a global result, applicable to  the whole piecewise differentiable curve. In
this connection it is also worth observing that, although lacking full generality, the differentiable case is  interesting on its own: for~example, most
physical actuators, whose constraints are obtained as solutions of differential equations, belong to this type of  context.\vspace{4pt}

In any geometric theory, covariance is of course  a key point. This aspect has already been taken care of in \cite{mbp1}, where the introduction of a transport
law for vertical vector fields along $\@\g\@$ yielded a covariant characterization of the ``\@true\@'' degrees of  freedom of the system.
Unhappily, in the standard approach, the local representation of the second variation involves non--tensorial terms.  In order to overcome this aspect we shall
develop an ``\@adaptation\@'' technique consisting in replacing the original Lagrangian by a gauge equivalent one  characterized by a suitably behaved
\emph{essential Hessian\/} along the given extremal.\vspace{4pt}

The paper is organized as follows.\vspace{4pt}

On the first instance, Section $1$ provides a brief summary of the contents of \cite{mbp1}\@:
the  geometric set-up for the formulation of the variational problem is outlined, the intrinsic characterization of  abnormality of evolutions is given and the
familiar Pontryagin equations for the extremal evolutions are drawn.\vspace{4pt}

Section $2$ represents the core of the paper. Here, taking a given extremal curve $\g$ into account, both necessary and sufficient conditions for minimality
are established by analysing the second variation of $\I\@[\g]\@$. To improve readability, some technicalities are deferred to Appendix \ref{SA}, notably a smoothing
theorem, extending to the non--holonomic context a well-known result in the holonomic framework.\vspace{4pt}

Finally, Section $3$ provides a plain geometric picture of the achieved results by reinterpreting them in terms of the  extremals of the \emph{accessory
variational problem}, commonly known as the \emph{Jacobi vector fields}.

\section{Geometric setup}\label{S1}
\subsection{Preliminaries}\label{S1.1}
In this Section we present a brief review of the geometric tools involved in the subsequent discussion. All results are stated without proof. The reader is
referred to \cite{mbp1,BLP} and references therein for a thorough description of the subject.

Throughout the paper, we shall freely use the language and methods of  differential geometry \cite{Sternberg,Warner}. The terminology will be partly borrowed
from classical non--holonomic Mechanics \cite{MP2,Saunders, Pom,DeLeon}.

Calculus of Variations has a very wide literature. Here we mention only some classical books \cite{Bliss, Lanczos, Gelfand, Giaquinta, Rund} along with those
more oriented to the issues arising from the presence of constraints \cite{Hestenes, Sagan, Pontryagin, Young} and those characterized by a geometric approach
\cite{Griffiths, Sussmann, Respondek, Montgomery, Agrachev1}.

\medskip \noindent
\textbf{(\/i\/)} \,Let $\@\V\rarw{t}\R\@$ denote a fibre bundle over the real line, henceforth called the \emph{event space\/}, and referred to local fibred
coordinates $\@t,q^1,\dots,q^n$\vspace{1pt}.

Every section $\@\g\colon\R\to\V\,$ is interpreted as the evolution, parameterized in terms of the independent variable $\/t\/$, of an abstract system $\@\B\/$
with a finite number of degrees of freedom. The first jet--bundle $\@\j\V\rarw{\pi}\V\@$, referred to local jet--coordinates $\@t,q^i,\dot q^i\/$, is called
the \emph{velocity space\/}. The first jet--extension of $\@\g\?$ is denoted by $\@\j\g\colon\R\to\j\V\@$\vspace{1pt}.

The presence of differentiable constraints is accounted for by a commutative diagram of the form\vspace{2pt}
\begin{subequations}\label{1.1}
\begin{equation}
\begin{CD}
\A          @>i>>       \j\V            \\
@V{\pi}VV                @VV{\pi}V      \\
\V          @=          \V
\end{CD}
\end{equation}

\noindent where\@:
\begin{itemize}
\item
$\@\A\rarw{\pi\,}{\V}\,$ is a fibre bundle, representing the totality of \emph{admissible velocities\/};
\smallskip
\item
the map $\@\A\rarw{i\,}{\j\V}\,$ is an imbedding;
\smallskip
\item
a section $\@\g\colon\R\to\V\@$ is \emph{admissible} if and only if its first jet--extension $\@\j\g\@$ factors through $\@\A\@$, i.e.~if and only if there
exists a section $\@\h\g\colon\R\to\A\,$ satisfying $\@\j\g=i\cdot\h\g\@$. Under the stated circumstance the section $\@\h\g\@$, commonly referred to as the
\emph{lift\/} of $\@\g\@$, is called an admissible section of $\@\A\@$.
\end{itemize}

\medskip \noindent
Referring the submanifold $\@\A\@$ to fibred local coordinates $\/t,\@q^1\!\And q^n\!,z^1\!\And z^r$,
the imbedding $\@i\colon\A\to\j\V\@$ is locally represented as
\begin{equation}
\dot q^i = \psi^i\/(t,q^1\!\And q^n\!,z^1\!\And z^r) \qquad\; i=1,\ldots,n\@,
\end{equation}
\end{subequations}
while the admissibility condition for a section $\@\h\g:q^i=q^i\/(t)\@,\,z^A=z^A\/(t)\@$ reads
\begin{equation*}
\d q^i/d t \,=\, \psi^i\left(t,q^1\/(t),\ldots,q^n\/(t),z^1\/(t),\ldots,z^r\/(t)\right)\@.\vspace{2pt}
\end{equation*}

Every section $\@\s\colon\V\to\A\@$ is called a \emph{control} for the system. The term is intuitively clear: assigning the section
$\@\s\colon\,z^A=z^A\/(t,q^1\And q^n)\@$ does in fact determine the evolution of $\@\B\/$ from given initial data through the solution of the first order
system of ordinary differential equations
\begin{equation*}
\d q^i/d t\,=\,\psi^i\big(\?t,q^1\!\And q^n\!,z^1\/(t,q^1\!\And q^n),\ldots,z^r\/(t,q^1\!\And q^n)\big)
\end{equation*}

\smallskip
\subsection{Geometry of the velocity space}\label{S1.2}
Given the event space $\@\V\@$, we denote by $\@V\/(\V)\@$\vspace{.4pt} the vertical bundle associated with the fibration $\@\V\to\R\@$ and by $\@V^*\/(\V)\@$
the corresponding \emph{dual bundle\/}.

By definition, $\@V^*\/(\V)\@$ is canonically isomorphic to the quotient of the cotangent bundle $T^*\/(\V)$ by the equivalence relation
\begin{equation}\label{1.2}
\s\sim\s'\;\Longleftrightarrow\;\left\{
\begin{aligned}
 & \pi\/(\s)=\pi\/(\s')\\[2pt]
 & \s-\s'\;\propto\; d\/t_{\,|\pi\/(\s)}
\end{aligned}
\right.
\end{equation}

\vskip1pt
For simplicity, we preserve the notation $\@\langle\;\;,\;\,\rangle\@$ for the pairing between $\@V(\V)\@$ and $\@V^*(\V)\@$.
The elements of $V^*\/(\V)\@$ are called the \emph{virtual $1$--forms\/} over $\V\@$.

For each $g\in\F\?(\V)$, the section $\delta\?g\colon\V\to V^*\/(\V)$ given by $\delta\?g_{|x}:=[\?d\/g_{|x}\?]$ is called the \emph{virtual differential\/}
of $\@g\@$.
Every element belonging to the tensor algebra generated by $\@V\/(\V)\@$ and $\@V^*\/(\V)\@$ is called a \emph{virtual tensor\/} over $\@\V\@$.

Every local coordinate system $t,q^i\/$ in $\V\@$ induces fibred coordinates $t,q^i,p_{\?i}\@$ in $\@V^*\/(\V)\,$, uniquely defined by the condition
$\,\l=p_{\?i}\/(\l)\,\delta\?q^i{}_{|\pi\/(\l)}\;\forall\,\l\in V^*(\V)\@$ and obeying the transformation laws
\begin{equation*}
\ovl t\@=\@t\@+\@c\,,\qquad \ovl q\@^i\@=\@\ovl q\@^i\/(t,q^1\!\And q^n)\,,\qquad \ovl p_{\?i}\@=\@p_{\?k}\,\de\?q^k/de{\@\ovl q\@^i}\,.
\end{equation*}

\medskip \noindent
\textbf{(\/ii\/)} \,\,The pull--back of $\@V^*\/(\V)\@$ through the map $\@\j\V\xrightarrow{\pi\,}\V\@$ determines a
\linebreak 
$\@(3\?n+1)\@$--dimensional manifold
$\@\C\@(\j\V)\@$, called the {\it contact bundle\/}. The latter is at the same time a vector bundle over $\/\j\V\@$, isomorphic to the subbundle of the
cotangent space $\@T^*\/(\j\V)\@$ locally generated by the $1$--forms $\@d\/q^i-\dot q^i\/d\/t\@$, and an affine bundle over $\@V^*\/(\V)\@$. The corresponding
projections are respectively denoted by $\@\C\/(\?\j\V)\rarw\k\j\V\@$ and $\@C\/(\?\j\V)\rarw\z V^*\/(\V)\@$.

\noindent
We shall refer $\@\C\@(\j\V)\?$ to coordinates $\@t,q^i,\dot q^i,p_{\?i}\@$ according to the prescription
\begin{equation}\label{1.3}
\s\@=\@p_{\?i}\/(\s)\@(d\/q^i-\@\dot q^i\@d\/t\?)_{\pi\/(\s)}\qquad\;\forall\;\s\in\C\@(\j\V)
\end{equation}
Every~$\@\s\in\C\/(\j\V)\@$ will be called a \emph{contact $1$--form\/} over $\@\j\V\@$.

Finally, we recall that, as a byproduct of the duality between $\@V\/(\V)\@$ and $\@V^*\/(\V)\@$, the manifold $\@\C\/(\?\j\V)\@$ is endowed with a
distinguished linear differential form, called the {\em Liouville $1$--form\/}, locally expressed as
\begin{equation*}
\Th\,=\,p_{\?i}\!\left(d\/q^i-\dot q^i\@d\/t\right).
\end{equation*}

\medskip\noindent
\textbf{(\/iii\/)} \,The restriction of $\@\C\/(\j\V)\@$ to the submanifold $\@\A\rarw i\j\V\@$ gives rise to a vector bundle $\C\/(\A)\to\A\@$, called the
contact bundle over $\@\A\@$. The situation is summarized into the commutative diagram\vspace{3pt}
\begin{equation*}
\begin{CD}
\C\/(\A)        @>\h\imath>>            \C\/(\j\V)          @>\k>>              V^*\/(\V)       \\
@V{\z}VV                                 @VV{\z}V                               @VVV            \\
\A              @>i>>                     \j\V                @>\pi>>            \V             \\
@V{\pi}VV                               @VV{\pi}V                                 @|            \\
\V                  @=                     \V                  @=                \V
\end{CD}\vspace{10pt}
\end{equation*}

According to the latter, $\@\C\/(\A)\@$ is canonically isomorphic to the pull--back of the bundle $\@V^*\/(\V)\to\V\@$ through the fibration $\@\A\to\V\@$.
Furthermore, the imbedding $\@\h\imath\colon\C\/(\A)\to\C\/(\j\V)\@$ endows the manifold $\@\C\/(\A)\@$ with a distinguished $1$--form
$\@\tilde\Th:=\h\imath\@^*\/(\Th)\@$, called the Liouville $1$--form of $\@\C\/(\A)\@$.

Referring $\@\C\/(\A)\@$ to fibre coordinates $\@t,q^i,z^A,p_i\@$, related in an obvious way to the coordinates in $\@\A\@$ and in $\@V^*\/(\V)\@$, we have the
representation
\begin{equation}\label{1.4}
\tilde\Th\,=\,p_{\?i}\@\big(\@d\/q^{\?i}\@-\@\psi^{\?i}\@d\/t\@\big)\,:=\,p_i\,\@\wt^{\?i}\@.
\end{equation}

\begin{remark}\label{Rem1.1}
According to the stated definition, the contact bundle $\@\C\/(\A)\@$ coincides with the subbundle of $\@T^*\/(\A)\@$ locally spanned by the $1$--forms
$\wt^{\?i}$. Exactly as in eq.~(\ref{1.3}), this property is made explicit by the representation
\begin{equation*}
\s\@=\@p_{\?i}\/(\s)\,\wt^{\?i}{}_{|\@\pi\/(\s)}\qquad\;\forall\;\s\in\C\@(\A)
\end{equation*}

In particular, the zero section $\@\O\colon\A\to\C\/(\A)\@$ has an intrinsic meaning: the image $\@\O\?(\A)\@$ is therefore a distinguished submanifold of
$\@\C\/(\A)\@$, diffeomorphic to $\/\A\@$.

Due to this fact, for each $\@z\in\A\@$, the tangent space $\@T_{\?\O\?(z)}\/(\C\/(\A))\@$ admits a direct sum decomposition of the form
\begin{equation*}
T_{\?\O\?(z)}\/(\C\/(\A))\,=\,(\O\!\@_z){\plus02}_*\@T_z\/(\A)\@\oplus\@V_{\?\O\?(z)}\/(\C\/(\A))\@,
\end{equation*}
$\@V_{\?\O\?(z)}\/(\C\/(\A))\@$ denoting the vertical space relative to the fibration $\@\C\/(\A)\to\A\@$ at the point $\@\O\?(z)\@$.

On the other hand, the differential of the Liouville $1$--form (\ref{1.4}) maps every vertical vector
$\@W=W_i\@\big(\de/de{p_{\?i}}\big){\!\@\plus05}_{\O\/(z)}\@$ at $\@\O\?(z)\@$ into the (the pull--back of) a contact 1--form at $\@z\@$ according to the
relation $\@W\mapsto W\interior d\?\tilde\Th\@ =\@W_i\,\wt^{\?i}{}_{|\?\O\?(z)}\@$.

Summing up, we conclude that, for all $\@z\in\A\@$, the tangent space $\@T_{\?\O\?(z)}\/(\C\/(\A))\@$ is canonically isomorphic to the direct sum
$\@T_z\/(\A)\@\oplus\@\C_z\/(\A)\@$.
\end{remark}
\subsection{Vector bundles along sections}\label{S1.3}
Given any admissible section $\@\g\colon\R\to\V\@$, let $\@\Vg\rarw{t}\R\@$\vspace{1pt} denote the bundle of vertical vectors along $\@\g\@$. Likewise, let
$\@\Ag\rarw{t}\R\@$ denote the totality of vectors along the lift $\@\h\g\colon\R\to\A\@$, annihilating the $1$--form $\@(d\/t\?)_{\h\g}\@$.

We adopt fibred coordinates $\?t,u^i\@$ in $\@\Vg\@$ and $\?t,u^i\!,v^A\@$ in $\@\Ag\@$, according to the prescriptions
\begin{subequations}\label{1.5}
\begin{alignat}{2}
& X\,=\,u^i\/(X)\bigg(\de/de{q^i}\bigg)_{\!\g\/(t\/(X))}         &&\forall\;X\in\Vg\,,\hskip.6cm                  \\[4pt]
& \h X\,=\,u^i\/(\h X)\bigg(\de/de{q^i}\bigg)_{\!\h\g\/(t\/(\h X))}\!\!+\,v^A\/(\h X)\bigg(\de/de{z^A}\bigg)_{\!\h\g\/(t\/(\h X))}
\qquad\;&&\forall\;\h X\in\Ag\,.
\end{alignat}
\end{subequations}

\smallskip
As shown in \cite{mbp1}, the first jet--bundle $\@\j\Vg\@$ is canonically isomorphic to the space of vectors along $\@\j\g\@$ annihilating the $1$--form
$\?dt\/$. Referring $\@\j\Vg\@$ to jet--coordinates $\@t\@,u^i,\dot u^i\@$, we have therefore the representation
\begin{equation*}
Z\,=\,u^i\/(Z)\bigg(\de/de{q^i}\bigg)_{\!\j\g\/(t\/(Z))}+\;\dot u^i\/(Z)\bigg(\de/de{\dot q^i}\bigg)_{\!\j\g\/(t\/(Z))}
\qquad\;\forall\;Z\in\j\Vg\,.
\end{equation*}

The push--forward of the imbedding $\@\A\rarw{i}{\j\V}\@$, restricted to the subspace $\@\Ag\subset T\?(\A)\@$, makes the latter into a subbundle of
$\@\j{\Vg}\@$. This gives rise to a fibred morphism
\begin{subequations}\label{1.6}
\begin{equation}
\begin{CD}
\Ag          @>i_*>>       \j\Vg            \\
@V{\pi_*}VV                @VV{\pi_*}V      \\
\Vg\;\;      @=             \Vg
\end{CD}\vspace{5pt}
\end{equation}
expressed in coordinates as\vspace{5pt}
\begin{equation}
\dot u^i=\left(\de\psi^i/de{q^k}\right)_{\!\h\g}\@u^k + \left(\de\psi^i/de{z^A}\right)_{\!\h\g}\@v^A\@.
\end{equation}
\end{subequations}

\smallskip
The kernel of the projection $\@\Ag\rarw{\pi_*}\Vg\@$, denoted by $\@V\/(\h\g)\@$, is called the \emph{vertical subbundle} along $\@\h\g\@$.

\medskip\noindent
\textbf{(\/ii\/)} \,The restriction of the space $\@V^*\/(\V)\@$ to the curve $\@\g\@$ determines a vector bundle $\@V^*\/(\g)\rarw{t}\R\@$, \emph{dual\/} to
the vertical bundle $\@\Vg\@$.

The elements of $V^*\/(\g)\@$ are called the \emph{virtual $1$--forms\/} along $\g\@$. The elements of the tensor algebra generated by $\@\Vg\@$ and
$\@V^*\/(\g)\@$ are called the \emph{virtual tensors\/} along $\@\g\@$.

As implicit in its definition, a virtual $1$--form at a point  $\@\g\/(t)\@$ is not a $1$--form in the ordinary sense, but an equivalence class of $\@1$--forms
under the relation (\ref{1.2}). Preserving the notation $\@\delta\?q^i\@$ for the equivalence class $\@[\@d\/q^i\@]\@$, every virtual tensor field
$\@W\colon\R\to \Vg\otimes_R V^*\/(\g)\otimes_R\cdots\,$ is locally represented as
\begin{equation*}
W\@=\,W^{\@i}{}_{j\@\cdots}\/(t)\,\bigg(\de/de{q^i}\bigg)_{\!\g}\!\otimes\wg j\otimes\cdots\@.
\end{equation*}

Every local coordinate system $t,q^i\/$ in $\V\@$ induces fibred coordinates $t,q^i,\pi_i\@$ in $\@V^*\/(\g)\@$, uniquely defined by the relation
\begin{equation*}
\l\@=\@\biggl<\@\l\,\@,\,\biggl(\de/de{q^i}\biggr)_{\!\g}\,\@\biggr>\;\wg i\@:=\,\pi_i\/(\l)\,\wg i
\qquad\quad \forall\,\l\in V^*\/(\g)\@.
\end{equation*}
\subsection{Admissible deformations}\label{S1.4}
In the presence of constraints, a deformation\linebreak $\@\g\?_\xi\colon\R\to\V\,$ of an admissible section $\@\g\@$ --- and, likewise, a deformation
$\@\h\g\?_\xi\colon\R\to\A\,$ of the lift $\@\h\g\@$ --- are called \emph{admissible\/} if and only if all sections $\@\g\?_\xi\@,\@\h\g\?_\xi\@$,
$\,\xi\in(\/-\/\eps,\eps\@)\@$ are admissible.

In coordinates, the admissible deformations of $\?\h\g\@$ are locally represented by equations of the form
\begin{equation*}
\h\g\?_\xi\@:\qquad q^i=\vphi^i\/(\xi,t)\,,\quad z^A=\z^A\/(\xi,t)\@,
\end{equation*}
subject to the conditions
\begin{subequations}\label{1.7}
\begin{align}
 & \vphi^i\/(0,t)\@=\@q^i\/(t)\,,\quad \z^A\/(0,t)\@=\@z^A\/(t)\@,                  \\[3pt]
 &\de\vphi^i/de t\,=\,\psi^i\/\big(t,\vphi^i\/(\xi,t),\z^A\/(\xi,t)\big)\@.
\end{align}
\end{subequations}

\smallskip
Setting $\@X^i\/(t):=\big(\de\@\vphi^i/de{\?\xi\;}\big)_{\xi=0}\,\@,\,X^A\/(t):=\big(\de\@\z\/^A/de{\?\xi\;}\big)_{\xi=0}\,$\vspace{1pt}, the infinitesimal
deformation tangent to $\@\h\g\?_\xi\,$ is the section $\@\h X\colon\R\to\A\?(\h\g)\@$ locally expressed as
\begin{equation*}
\h X\,=\,X^i\/(t)\left(\de/de{q^i}\right)_{\!\h\g}\,+\,X^A\/(t)\left(\de/de{z^A}\right)_{\!\h\g}\@,
\end{equation*}
while the admissibility condition (\ref{1.7}\@b) is reflected into the \emph{variational equation\/}
\begin{equation}\label{1.8}
\d X^i/d t\,=\,\left(\de\@\psi^i/de{q^k}\right)_{\!\h\g}X^k\,+\,\left(\de\@\psi^i/de{z^A}\right)_{\!\h\g}X^A\@.
\end{equation}
The infinitesimal deformation tangent to the projection $\@\g\?_\xi=\pi\cdot\h\g\?_\xi\,$ is similarly defined as the section $\@X\colon\R\to\Vg\@$ locally
expressed by
\begin{equation*}
X=\pi_*\@\h X\,=\,X^i\/(t)\@\left(\de/de{q^i}\right)_{\!\h\g}\@.
\end{equation*}

\smallskip
The previous arguments point out a complete symmetry between the roles of diagram (\ref{1.1}\@a) in the study of the admissible \emph{evolutions\/} and of
diagram (\ref{1.6}\@a) in the study of the admissible infinitesimal \emph{deformations\/}, thus enforcing the intuitive viewpoint that the latter context is
essentially a ``linearized counterpart'' of the former one.
\medskip
\subsection{Infinitesimal controls}\label{S1.5}
Given an admissible evolution $\@\g\colon\R\to\V\@$, an \emph{infinitesimal control} along $\@\g\@$ is a linear section $\@h\colon\Vg\to\Ag\,$. The image
$\@\Hg:=h(\Vg)\@$ is called the \emph{horizontal distribution} along $\@\h\g\@$ induced by $\@h\@$. Every section $\@\h X\colon\R\to\Ag\@$ satisfying
$\@\h X\/(t)\in\Hg\;\,\forall\,\@t\in\R\@$ is called a horizontal section of $\@\Ag\@$\vspace{1pt}.

In fibre coordinates an infinitesimal control is locally represented as\vspace{1pt}
\begin{equation*}
v^A\,=\,h_i^{\,A}\/(t)\, u^i\@.
\end{equation*}
while the associated horizontal distribution is locally spanned by the vector fields
\begin{equation}\label{1.9}
\DE_i\,:=\,h\@\bigg[\left(\de/de{q^i}\right)_{\!\g}\@\bigg]\,=\,\left(\de/de{q\?^i}\right)_{\!\h\g} + h\@_i{}^A\,\left(\de/de{z^A}\right)_{\!\h\g}\@.
\end{equation}

\smallskip
By means of $\@h\@$, every section $\@X\@=\@X^i\/(t)\,\big(\de/de{q^i}\big)_{\!\g}\@$ of $\@\Vg\@$  may be lifted to a horizontal section
$\@h\/(X)=X^i\,\DE_i\@$ of $\@\Ag\@$.\vspace{1pt}

More crucially, every section $\@\h X=X^i\/(t)\,\big(\de/de{q^i}\big)_{\h\g} + X^A\/(t)\,\big(\de/de{z^A}\big)_{\h\g}\@$\vspace{1pt} of~$\Ag\@$ may
be uniquely decomposed into the sum\vspace{1pt} of a horizontal part $\@\P_H\/(\h X)\@$ and a vertical part $\@\P_V\/(\h X)\@$, respectively defined by the
equations
\begin{subequations}\label{1.10}
\begin{align}
&\P_H\/(\h X)\,:=\,h\,\big(\pi_*\/(\h X)\big)\,=\,X^i\,\DE_i\@,                                         \\[4pt]
&\P_V\/(\h X)\,:=\,\h X\@-\@\P_H\/(\h X)\,=\,\big(\@X^A-X^i\@h\@_i{}^A\big)\@\bigg(\de/de{z^A}\bigg)_{\!\h\g}\@.
\end{align}
\end{subequations}

\medskip\noindent
\textbf{(\/ii\/)} \,A section $X\colon\R\to\Vg$\vspace{.7pt} is said to be $h$--transported along $\@\g\@$ if the horizontal lift $\?h\/(X)\@$ is an admissible
infinitesimal deformation of $\@\h\g\@$, i.e.~if it satisfies the condition\linebreak 
$\@i_*\cdot h\/(X)=\j X\@$.
In coordinates, this amounts to the requirement
\begin{equation*}
\d X^i/d t\@=\left[\bigg(\de\psi^i/de{q^k}\bigg)_{\!\h\g}\, +\,h_k{}^A\,\bigg(\de\psi^i/de{z^A}\bigg)_{\!\h\g}\right]\@X^k\@ = \@X^k\,\DE_k\@\psi^i\@.
\end{equation*}

By Cauchy theorem, the $h$--transported sections form an $n$--dimensional vector space $\@V_h\@$, isomorphic to each fibre $\@\Vg\?_{|\@t}\@$. This provides a
\emph{trivialization\/} of the vector bundle $\@\Vg\to\R\@$, summarized into the identification $\@\Vg\simeq\R\times V_h\@$.

The dual space $\@{V_h}^{\!*}\@$ gives rise to an analogous trivialization $\@V^*\/(\g)\simeq \R\times {V_h}^{\!*}\@$.

\medskip\noindent
\textbf{(\/iii\/)} \,The notion of $h$--transport induces an \emph{absolute time derivative\/} for vertical vector fields along $\@\g\@$. The operation is
naturally extended to a derivation of the algebra of virtual tensor fields along $\@\g\@$, commuting with contractions.

In coordinates, introducing the \emph{temporal connection coefficients\/}
\begin{equation}\label{1.11}
\tau\?_k{}^i\,:=\,-\,\DE_k\!\left(\psi^i\right)=\,-\left(\de\psi^i/de{q\?^k}\right)_{\!\h\g}\@-\@
h\@_k{}^A\/\left(\de\psi^i/de{z^A}\right)_{\!\h\g}
\end{equation}
and adopting the notation
\begin{equation*}
\D/Dt\biggl[\@W^{\@i}{}_{j\,\cdots}\/(t)\,\biggl(\de/de{q^i}\biggr)_{\!\g}\!\otimes\@\wg{j}\otimes\cdots\,\biggr]:=\,
\D\@W^{\@i}{}_{j\,\cdots}/Dt\,\biggl(\de/de{q^i}\biggr)_{\!\g}\!\otimes\@\wg{j}\otimes\cdots\;\;
\end{equation*}
we have the representation
\begin{equation}\label{1.12}
\D\@W^{\@i}{}_{j\,\cdots}/Dt\;=\;\d\@W^{\@i}{}_{j\,\cdots}/dt\;+\;\tau\?_k{}^i\;W^{\@k}{}_{j\,\cdots}\;-
\;\tau\?_j{}^k\;W^{\@i}{}_{k\,\cdots}\;+\;\cdots
\end{equation}

\smallskip
The algorithm\vspace{1pt} may be simplified referring both vector bundles $\@\Vg\/$, $\@V^*\/(\g)\@$ to \linebreak
$h$--transported dual bases
$\@e_\na=e_\na^{\;\,i}\big(\de/de{q^i}\big)_\g\,$, $\;e^\na=e^\na_{\;\,i}\,\wg{i}\,$.

Setting $\@W^\na{}_{\nb\,\cdots}:= \big< W,\@e^\na\otimes e_\nb\otimes\cdots\,\big>\@=\@ W^{\@i}{}_{j\,\cdots\?}\; e^\na_{\;\,i}\,e_\nb^{\;\,j}\cdots\@$, we
have in fact the representations
\begin{equation*}
W\@=\@W^\na{}_{\nb\,\cdots}\,e_\na\otimes e^\nb\otimes\cdots\@,\quad\,
\D\@W/Dt\,=\,\d\@W^\na{}_{\nb\,\cdots}/dt\;\@e_\na\otimes e^\nb\otimes\cdots\@.
\end{equation*}

\medskip\noindent
\textbf{(\/iv\/)} \,In view of eqs.~\!(\ref{1.10}\@a,\@b), every infinitesimal deformation $\@\h X\@$\vspace{1pt} of the section $\@\h\g\@$ admits a unique
representation of the form $\@\h X=h\/(X)+U$, where:
\begin{itemize}
\item
$\@X\@=\@\pi_*\@\h X\@:=\@X^i\@\big(\de/de{q^i}\big)_\g\@$ is a vertical field along $\@\g\@$, namely the (unique) infinitesimal deformation of $\@\g\@$
lifting to $\@\h X\@$;
\smallskip
\item
$\@U\@\@=\@\P_V\/(\h X)\@:=\@U^A\@\big(\de/de{z^A}\big)_{\h\g}\@$ is a vertical vector field along $\@\h\g\@$.
\end{itemize}
In terms of this decomposition, the variational equation (\ref{1.8}) takes the form
\begin{equation*}
\d X^i/dt\,=\,X^k\bigg(\de\psi^i/de{q^k}\bigg)_{\!\h\g}+\,
\bigg(\de\psi^i/de{z^A}\bigg)_{\!\h\g}\left(X^k\@h_k{}^A\@+\@U^A\right)\@.
\end{equation*}
On account of eqs.~\!(\ref{1.11}), (\ref{1.12}), the latter is more conveniently written as
\begin{subequations}\label{1.13}
\begin{equation}
\D X^i/D t\,=\,\bigg(\de\psi^i/de{z^A}\bigg)_{\!\h\g}U^A
\end{equation}
or also, in $h$--transported bases,
\begin{equation}
\d X^\na /d t = e^\na_{\;\,i}\bigg(\de\psi^i/de{z^A}\bigg)_{\!\h\g}U^A\@.
\end{equation}
\end{subequations}

Setting $\@\psi^\na_{\,A}:=e^\na_{\;\,i}\big(\de\psi^i/de{z^A}\big)_{\!\h\g}\@$ we conclude that every infinitesimal deformation is determined, up to initial
data, by the knowledge of a vertical vector field along $\@\h\g\@$, through the equation
\begin{equation}\label{1.14}
X^i\/(t)\@=\@X^\na\/(t)\,e_\na^{\;\,i}\/(t)=\biggl(X^\na\/(t_0)\@+\@\int_{t_0}^{t}\@\psi^\na_{\,A}\@U^A\,d\tau\biggr)\,e_\na^{\;\,i}\/(t)\@.
\end{equation}
\smallskip
\subsection{Extremals}\label{S1.6}
Let $\@\Lagr\in\F(\A)\@$ denote a differentiable function on the manifold $\@\A\@$, henceforth called the \emph{Lagrangian}. Constrained calculus of variations
deals with the study of the extremals of the functional
\begin{equation*}
\I\?[\g]:=\int_{\h\g}\Lagr\/(t, q^1, \ldots, q^n, z^1, \ldots, z^r)\, d\/t
\end{equation*}
among all admissible closed arcs $\@\g\colon[\@t_0,t_1]\to\V\@$ connecting two given configurations\@
\footnote
{\@We recall that an admissible closed arc $\@\g\colon[\@t_0,t_1]\to\V\@$ is the restriction to $[\@t_0, t_1]\@$ of an admissible section $\g \colon I\to\V\@$
defined on some open interval $\@I\supset [\@t_0,t_1]\@$.}. We shall refer to this as \emph{the control problem\/}.

In this Section we review the stationarity conditions for $\@\I\?[\g]\@$. Once again, all results are stated without proof, referring to \cite{mbp1} for the
full argumentation.

\medskip\noindent
\textbf{(\/i\/)} \,Given an admissible section $\@\g\colon[\@t_0,t_1]\to\V\@$, let $\@\mathfrak V\@$ denote the infinite dimensional vector space formed by the
totality of vertical vector fields $\@U=U^A\@\big(\de/de{z^A}\big)_{\h\g}\@$ along $\@\h\g\@$. Consider the linear map $\@\Upsilon:\mathfrak
V\to\Vg\?_{|\@t_1}\@$ defined by the equation
\begin{equation}\label{1.15}
\Upsilon\?(U)\@=\@\left(\,\int_{t_0}^{t_1}e^\na_{\;\,i}\bigg(\de\psi^i/de{z^A}\bigg)_{\!\h\g} U^A\,d\/t\right)e_{(a)}\@.
\end{equation}

Setting $X^\na\/(\@t_0)=0$ in eq.~\!(\ref{1.14}), it is readily seen that the subspace $\ker\@(\Upsilon)\subset\mathfrak V\@$
\linebreak
is in 1--1 correspondence with the vector space formed by the totality of the admissible infinitesimal deformations vanishing at the endpoints of~$\g\@$.

The co--dimension $\@n(\g)\@$ of the image space $\@\Upsilon\?(\mathfrak V)\subset\Vg\?_{|\@t_1}\@$ expresses the \emph{abnormality index\/} of $\@\g\@$.
Depending on the value of the latter, the admissible sections are classified into \emph{normal\/}, when $\@n(\g)=0\@$, i.e.~\!when the map (\ref{1.15}) is
surjective, and \emph{abnormal\/}, when $\@n(\g)>0\@$.
A section $\@\g\@$ is called \emph{locally normal\/} if its restriction to any closed subinterval $[\@a,b\@]\subseteq[\@t_0,t_1]$ is normal.

Concerning the evaluation of $\@n(\g)\@$ we have the result \cite{mbp1}\@:
\begin{subequations}\label{1.16}
\begin{proposition}
The annihilator $\@\big(\@\Upsilon\?(\mathfrak V)\@\big){}^0\subset\Vg\?_{|\@t_1}^*\@$ coincides with the totality of virtual \mbox{$1$--forms}
$\@\l=\l_{\?i}\/(t)\,\wg{i}\@$ satisfying the conditions
\begin{equation}
\d \l_i/d t\,+\,\l_k\@\bigg(\de\psi^k/de{q^i}\bigg)_{\!\h\g}=\,0\,,\quad\;\l_i\@\bigg(\de\psi^i/de{z^A}\bigg)_{\!\h\g}=\,0\@.
\end{equation}
\end{proposition}

\begin{remark}\label{Rem1.2}
On account of eqs.~\!(\ref{1.11}), (\ref{1.12}), given any infinitesimal control along $\@\g\@$ and denoting by $\@\D/Dt\@$ the corresponding absolute time
derivative, eqs.~\!(\ref{1.16}\@a) are mathematically equivalent to the system
\begin{equation}
\D\l/Dt\,=\,0\,,\qquad\; \l_i\@\bigg(\de\psi^i/de{z^A}\bigg)_{\!\h\g}=\,0\@.
\end{equation}
In particular, a section $\@\g\@$ is locally normal if and only if eqs.~\!(\ref{1.16}) do not admit any solution other than the trivial one $\@\l=0\@$ on every
subinterval $[\@a,b\@]\subseteq[\@t_0,t_1]\@$.
\end{remark}
\end{subequations}

\begin{remark}\label{Rem1.3}
As it is clear from the definition, local normality implies normality.
\linebreak
The converse is generally false, as shown by the following example: assume an imbedding $\@\A\rarw{i}\j\V\@$ locally described by the equations
\begin{equation*}
\left\{
\begin{alignedat}{2}
  & \dot q^A&&=z^A\qquad\qquad A=1\And n-1  \\
  & \dot q^n&&=f\/(t)\,z^1
\end{alignedat}
\right.
\end{equation*}
with $\,f(t)=\exp\?(-\@\nicefrac1{\plus60 t^2})\,$ for $\@t<0\,$ \vspace{.8pt} and $\,f(t)=0\,$ for $\@t\geq 0\@$.

\noindent
Along any admissible section $\@\g\colon[\@t_0,t_1]\to\V\@$ eqs.~\!(\ref{1.16}\@a) take the form
\begin{equation}\label{1.17}
\d\l_i/d t\,=\,0\,,\qquad\;\l_1\@+\@\l_n\@f\/(t)\,=\,0\,,\qquad\l_2\,=\,\cdots\,=\,\l_{n-1}\,=\,0\,.
\end{equation}
In particular, if $\@t_0<0<t_1\@$ we conclude that:
\begin{itemize}
\item
$\@\g\@$ is normal, since eqs.~\!(\ref{1.17}) do not admit any non-zero solution for $\@t\in[\@t_0,t_1\?]\@$;
\item
$\@\g\@$ is not locally normal, since eqs.~\!(\ref{1.17}) admit the solution $\@\l_1=\cdots=\l_{n-1}=0\@$, $\@\l_n=\const\@$ along any subinterval
$[\@a,b\@]\subseteq[\@0,t_1]\@$.
\end{itemize}
\end{remark}

\smallskip\noindent
\textbf{(\/ii\/)} \,The abnormality index $\@n(\g)\@$ may be alternatively characterized in terms of the Liouville $1$-form (\ref{1.4}). The latter embodies
the manifold $\@\C\/(\A)\@$ with the intrinsic action functional
\begin{equation}\label{1.18}
\I_0\@[\?\gt]\@:=\,\int_\gt\@\tilde\Th\,=\,\int_{t_0}^{t_1} \!p_i\left(\d q^i/dt\@-\@\psi^i
\right)\,dt\@.
\end{equation}
The resulting extremals, locally expressed as $\@\gt\colon q^i=q^i\/(t)\@,\,z^A=z^A\/(t)\@,\,p_i=p_i\/(t)\@$, are solutions of the Euler--Lagrange equations
\begin{equation}\label{1.19}
\d q^i/d t\,=\,\psi^i\/(t\?,\@q^i,\@z^A)\,, \qquad\d p_i/d t\,+\,\de\psi^k/de{q^i}\,p_k\,=\,0\,, \qquad p_i\,\de\psi^i/de{z^A}\,=\,0\@.
\end{equation}
From these, taking eqs.~\!(\ref{1.16}) into account, we draw the following conclusions:
\begin{itemize}
\item
a section $\g\colon[\@t_0,t_1]\to\V$ is admissible if and only if the functional $\@\I_0\@[\?\gt]\@$ admits at least one extremal $\@\gt\@$ projecting onto
$\@\h\g\@$ i.e.~satisfying $\@\z\cdot\gt=\h\g\@$;

\vskip1pt
\item
the totality of extremals of $\@\I_0\@[\?\gt]\@$ projecting onto an admissible section $\@\g\@$ form a finite dimensional vector space, whose dimension
coincides with the abnormality index of $\@\g\@$;

\vskip1pt
\item
for each \emph{normal\/} $\@\g\@$, the unique extremal of $\@\I_0\@[\?\gt]\@$ projecting onto $\@\g\@$ is the composite section $\@\gt=\O\cdot\h\g\@$,
i.e.\;the image of the curve $\@\h\g\colon[\@t_0,t_1]\to\A\@$ through the null section $\@\O\colon\A\to\C\?(\A)\@$.
\end{itemize}

\medskip\noindent
\textbf{(\/iii\/)} \,Under suitable assumptions, the control problem may be converted into a free variational problem on the contact bundle $\@\C\/(\A)\@$.
This is achieved by lifting the Lagrangian $\@\Lagr\@$ to a $1$--form $\@\thL\@$ over $\@\C\/(\A)\@$ according to the prescription\@
\footnote%
{\@A deeper insight into the geometrical meaning of $\@\thL\@$ comes from the study of the gauge--theoretical structure of the control problem,
as developed in \cite{BLP}.}%
\footnote%
{\@For simplicity, we shall use the same symbols both for covariant fields and for their pull--back, namely we shall write $\@\psi^i\@$ for
$\@\z^*\/(\psi^i)\@,\,\@\wt^{\?i}\@$ for $\@\z^*\/(\wt^{\?i})\@$ etc.}:
\begin{equation*}
\thL\,:=\,\Lagr\,dt\@+\@\tilde\Th\,=\,(\Lagr-p_i\,\psi^i)\,dt\@+\@p_i\,dq^i\,:=\,-\,\Ham\,dt\@+\@p_i\,dq^i\@.
\end{equation*}
The function $\@\Ham:=p_i\,\psi^i-\@\Lagr\in\F\?(\C\/(\A))\@$\vspace{1pt} is called the \emph{Pontryagin Hamiltonian\/}. The Euler--Lagrange equations
associated with the action functional $\@\I\@[\gt]:=\int_\gt \thL\@$ take the form
\begin{equation}\label{1.20}
\d q^i/d t\,=\,\psi^i(t,q^i,z^A)\,, \qquad \d p_i/dt\,+\,\de\psi^k/de{q^i}\,p_k\,=\,\de \Lagr/de{q^i}\,,\qquad p_i\,\de\psi^i/de{z^A}\,=\,\de \Lagr/de{z^A}\,,
\end{equation}
i.e.\;they coincide with the \emph{Pontryagin equations\/} associated with the original control problem. Hence the result:
\begin{theorem}
An admissible section $\@\g\colon[\@t_0,t_1]\to\V\@$ is an extremal of the functional $\@\I\@[\g]\@$ if and only if there exists an extremal
$\@\gt\colon[\@t_0,t_1]\to
\C\/(\A)\@$ of $\@\I\@[\gt]\@$ projecting onto $\@\g\@$, i.e.~satisfying $\@\z\cdot\gt=\h\g\@$. In particular, whenever $\@\g\@$ is a \emph{normal\/}
extremal, there exists a unique such $\@\gt\@$.\@\footnotemark{}
\end{theorem}

\footnotetext%
{\plus80 \@Needless to say, this $\@\gt\@$ has nothing to do with the null section $\@\O\cdot\h\g\@$, unless the Lagrangian satisfies the conditions
$\@\Big(\de\Lagr/de{q^i}\Big)_{\h\g}=\@\Big(\de\Lagr/de{z^A}\Big)_{\h\g}=\@0\@$. We shall return on this point in Sec.\!~\ref{S2.1}.}

\medskip\noindent
\textbf{(\/iv\/)}\, A point $\@\vsigma\in \C\/(\A)\@$ is called \emph{regular} if and only if the equation
\begin{equation*}
p_i\,\de\psi^i/de{z^A}\,=\,\de \Lagr/de{z^A}\qquad\left(\;\Longleftrightarrow\;\de\Ham/de{z^A}\,=\,0\@\right)
\end{equation*}
can be uniquely solved for $\@z^1,\dots,z^r\@$ in a neighborhood of $\@\vsigma\@$, giving rise to local expressions of the form
\begin{equation}\label{1.21}
z^A\,=\,z^A\?(t,\?q^1,\@\ldots,\@q^n,\@p_1,\@\ldots,\@p_n)\@.
\end{equation}
A sufficient condition for this to happen is the validity of the condition
\begin{equation}\label{1.22}
\det\bigg(\@\sd\Ham/de z^A/de{z^B}\bigg)_{\!\vsigma}\@\neq\,0\@.
\end{equation}

In a neighborhood of each regular point, substituting eq.~\!(\ref{1.21}) into the first pair of equations (\ref{1.20}) and setting
$\@\H(t,q^i,p_i):=\Ham(\/t,q^i,p_i,z^A(t,q^i,p_i)\/)\@$\vspace{1pt} allows to cast the Pontryagin equations in the Hamiltonian form
\begin{equation*}
\d q^i/dt\,=\,\de \H/de{p_i}\,, \qquad \d p_i/dt\,=\,-\,\de \H/de{q^i}\@.
\end{equation*}

A normal extremal $\@\g\@$ of the functional $\@\I\?[\g]\@$ is called \emph{regular\/} if and only if the condition (\ref{1.22}) holds throughout the (unique)
extremal of $\@\I\?[\gt]\@$ projecting onto $\@\g\@$.
\section{The second variation of the action functional}\label{S2}
By definition, the (weak) extremals of the action functional $\@\I\?[\g]=\int_{\h\g}\@\Lagr\@\@dt\@$ are sections $\@\g\colon[t_0,t_1]\to\V\@$\vspace*{1.5pt}
characterized by the vanishing of the first derivative $\@\d\I\?[\g_\xi]/d\xi\@{\vert{11}5}_{\@\xi=0}\,$ for all admissible deformations
$\@\g_{\?\xi}\colon[t_0,t_1]\to\V\@$ with fixed endpoints.

\smallskip
We shall now deal with the problem of establishing both necessary and sufficient conditions for a given extremal to provide a weak local minimum for
$\@\I\?[\g]\@$.\vspace{1pt} To this end, we consider the second derivative $\@\frac{d^{\@2}\I\?[\g_\xi]}{d\?\xi^{\?2}}\@{\vert{11}5}_{\@\xi=0}\,$, commonly
known as the \emph{second variation} of the action functional at $\?\g\,$.\vspace{1pt}

Before getting to the heart of the matter, we first take the necessary steps in order to simplify the algorithm and ensure the tensorial character of the
results.
\subsection{Lagrangians ``adapted'' to a given extremal curve}\label{S2.1}
Given any function $\@S\@$\vspace{1pt} over $\@\V\@$, let $\?\dot S:=\de S/de t+\de S/de{q^i}\,\psi^i\in\F\/(\A)\@$ denote its \emph{symbolic time derivative\/}\@
\footnote%
{\@Depending on the context, the same symbols will indicate the pull back of $\@S\@,\@\dot S\@$ through the various fibrations.}.
Any correspondence of the form $\@\Lagr\to\Lagr\?':=\Lagr-\dot S\@$ is called a \emph{gauge transformation\@}. Two Lagrangians $\@\Lagr\@$ and $\@\Lagr\?'$
differing by a symbolic time derivative are said to be \emph{gauge--equivalent\/}.

As it is well known, gauge equivalent Lagrangians determine the same extremal in $\@\V\@$. Matters are slightly different when we lift the original problem to
a corresponding free one in $\@\C\?(\A)\@$, along the lines of \?Sec. \!\ref{S1.6}. On account of the relation
\begin{equation*}
\dot S\@d\/t\@=\@d\?S\@-\@\de S/de{q^i}\,\big(d\/q^i-\@\psi^i\?d\/t\big)\@,
\end{equation*}
it is in fact readily seen that the extremals of the functional $\@\int_{\tilde\g} \@\th_{\hskip-.5pt\Lagr^{\@\prime}}\@$ differ from those of
$\@\int_{\tilde\g}\@\th_{\hskip-.5pt\Lagr}\@$ by a translation
\begin{equation*}
p^{\?\prime}_i\/(t)\;\@=\;p_i\/(t)\@-\,\de S/de{q^i}\@\big(\@t,q^i\/(t)\?\big)
\end{equation*}
along the fibres of $\@\C\?(\A)\@$: the lift process $\@\g\to\tilde\g\@$ is not a gauge invariant operation, but explicitly depends on the choice of the
Lagrangian. In particular, when working with a given $\@\g\@$, we can always fix the gauge by requiring that the extremals of
$\@\int_{\tilde\g}\@\th_{\hskip-.5pt\Lagr\?'}\@$ projecting onto $\@\g\@$ be also extremals of the purely geometrical functional (\ref{1.18}) associated with
the Liouville $1$--form $\@\tilde\Th\@$
\footnote%
{\@When $\@\g\@$ is a \emph{normal\/} extremal, there exists a unique such $\@\gt\@$, identical to the composite section
$\@\O\cdot\h\g\colon[\@t_0,t_1]\to\C\?(\A)\@$.}.

In view of eqs.~\!(\ref{1.19}), (\ref{1.20}), a necessary and sufficient condition for this to happen is the validity of the relations
$\@\big(\de\Lagr\?'/de{q^k}\big)_{\h\g}\@=\big(\de\Lagr\?'/de{z^A}\big)_{\h\g}\@=\@0\@$, i.e.~the vanishing of $\@\h X\/(\Lagr\?')\@$ for all $\@\h X\in\Ag\@$.
Setting $\@\Lagr\?'=\Lagr-\dot S\@$, the stated condition is summarized into the system of differential equations
\begin{multline*}
0=\biggl(\de\Lagr/de{q^k}\biggr)_{\!\h\g}\@-\@\biggl(\sd S/de{q^k}/de t\biggr)_{\!\g}\@-\@
\biggl(\sd S/de{q^k}/de{q^r}\biggr)_{\!\g}\,\psi^r{\!\plus03}_{|\@\h\g}-\@
\biggl(\de S/de{q^r}\biggr)_{\!\g}\,\biggl(\de\psi^r/de{q^k}\biggr)_{\!\h\g}=                       \\[3pt]
=\@\biggl(\de\Lagr/de{q^k}\biggr)_{\!\h\g}-\@\d/dt\@\biggl(\de S/de{q^k}\biggr)_{\!\g}-
\biggl(\de S/de{q^r}\biggr)_{\!\g}\biggl(\de\psi^r/de{q^k}\biggr)_{\!\h\g}\@,
\end{multline*}
\begin{equation*}
0=\biggl(\de\Lagr/de{z^A}\biggr)_{\!\h\g}-
\biggl(\de S/de{q^r}\biggr)_{\!\g}\@\biggl(\de\psi^r/de{z^A}\biggr)_{\!\h\g}\@. \hskip5.8cm
\end{equation*}
for the unknown $\@S\/(t,q^1\And q^n)\@$. The general solution of the latter may be locally written as
\begin{equation}\label{2.1}
S\,=\,p_i\/(t)\,q^i+ h\/(t,q^1\And q^n)\@,
\end{equation}
with $\@p_i\/(t)\@$ satisfying the second pair of Pontryagin equations (\ref{1.20}) and with $\@h\@$ being any solution of the homogeneous equation
\vspace{2pt}$\@\big(\de h/de{q^k}\big){\plus04\!}_\g\!=0$.

Every Lagrangian $\@\Lagr\?'\@$ satisfying the stated requirements will be said to be \emph{adapted\/} to the extremal $\@\g\@$. By construction, the class of
adapted Lagrangians is closed under \emph{gauge transformations of the second kind\/}
\begin{equation}\label{2.2}
\Lagr\?''\, =\ \Lagr\?'-\@\dot g\@,
\end{equation}
with $\@g=g\/(t,q^1\And q^n)\@$ and $\@\big(\de g/de{q^k}\big){\plus04\!}_\g\!=0\@$. When $\@\g\@$ is a \emph{normal\/} extremal, each pair of adapted
Lagrangians is related by a transformation (\ref{2.2}).

\medskip\noindent
\textbf{(\/ii\/)} \,Given any differentiable function $\@f\@$ on a manifold $\@M\@$, let $\@(d\/f_{|x})^0\subset T_x\/(M)\@$ denote the annihilator of the
differential $\@d\/f_{|x}\@$\vspace{.6pt} at a point $\@x\in M\/$.
A straightforward check then shows that the (generally non covariant)\vspace{.6pt} correspondence $\@T_x\/(M)\times T_x\/(M)\to\R\@$\vspace{.6pt} expressed in
coordinates as $\@X,Y\to\big(\sd f/de{x^i}/de{x^j}\big)_{\!x}\,X^i\,Y^j\@$\vspace{1pt} induces a covariant bilinear functional
 $\@(d\/f_{|x})^0 \times
(d\/f_{|x})^0\to\R\@$.\vspace{.6pt} The latter, henceforth denoted by $\@\big(d^{\,2}f\big)_x\@$, will be called the \emph{essential Hessian\/} of $\@f\@$ at
$\@x\@$.

Let us now recall that, by definition, the class of Lagrangians $\Lagr\?'$ adapted to~$\g$ is characterized by the requirement
$\A\/(\h\g\/(t))\subset(d\?\Lagr\?'\plus0{.4}_{|\h\g\/(t)})^0\@$ $\forall\,t\in[\?t_0\?,\?t_1\?]\@$.\vspace{.5pt}
Every such $\Lagr\?'$ determines therefore a symmetric bilinear functional $\big(d^{\,2}\Lagr\?'\big){}_{\@\h\g}\@$\vspace{1pt} along $\@\h\g\@$, whose action
on an arbitrary pair of vectors $\@\h X=X^i\,\big(\de/de{q^i}\big)_{\!\h\g\/(t)}\!+\@X^A\,\big(\de/de{z^A}\big)_{\!\h\g\/(t)}\@$\vspace{1pt}, $\@\h
Y=Y^i\,\big(\de/de{q^i}\big)_{\!\h\g\/(t)}\!+\@ Y^A\,\big(\de/de{z^A}\big)_{\!\h\g\/(t)}\!\in\A\?(\h\g\/(t))\@$ is expressed by the relation
\begin{multline}\label{2.3}
\Big<\big(d^{\,2}\Lagr\?'\big)_{\@\h\g}\,,\,\h X\otimes \h Y\@\Big>\,=\,
\bigg(\sd\Lagr\?'/de q^i/de{q^j}\bigg)_{\!\h\g\/(t)}\,X^i\,Y^j\,+                                           \\
+\@\bigg(\sd\Lagr\?'/de q^i/de{z^A}\bigg)_{\!\h\g\/(t)}\,\big(X^i\,Y^A\@+Y^i\,X^A\big)\@+
\@\bigg(\sd\Lagr\?'/de z^A/de{z^B}\bigg)_{\!\h\g\/(t)}\,X^A\,Y^B\@.\hskip.6cm
\end{multline}

In a similar way, if $\@g\in\F\/(\V)\@$ is any function satisfying $\@\big(\de g/de{q^k}\big)_{\!\g}\!=0\@$, the essential Hessian
$\big(d^{\,2}g\big){\plus0{1.5}}_{\?\g}$ determines a bilinear functional on $\@\Vg \times \Vg\@$, i.e.\;a virtual tensor of rank~2 along $\@\g\@$, expressed in
coordinates as
\begin{equation}\label{2.4}
\big(d^{\,2}g\big)_{\?\g}=\biggl(\sd g/de q^i/de{q^j}\biggr)_{\!\g}\@\wg{i}\otimes\wg{j}\@.
\end{equation}
Conversely, every symmetric virtual tensor $\@C=C_{ij}\/(t)\,\wg{i}\!\@\otimes\@\wg{j}\@$ along $\@\g\@$ may be obtained as the essential Hessian of a function
$\?g\in\F\?(\V)\@$ defined in a neighborhood of $\@\g\@$ (for instance $\@g=\frac12\,C_{ij}\/(t)\?(q^i-q^i\/(t))(q^j-q^j\/(t))\@$).\vspace{3pt}

Under the assumption $\@\big(\de g/de{q^k}\big)_{\!\g}\!=0\@$, the essential Hessian $\@(d^{\,2}\dot g)_{\?\h\g}\@$ is easily recognized to define a symmetric
bilinear functional on $\@\Ag \times \Ag\@$. Preserving the notation $\@\big(\sd g/de q^i/de{q^j}\big)_{\!\g}=C_{ij}\/(t)\@$, a straightforward calculation provides
the evaluation
\begin{multline}\label{2.5}
\hskip-8pt\Big<(d^{\,2}\dot g)_{\?\h\g}\@,\@\h X\otimes\h Y\@\Big>\,=\,
\biggl(\sd\dot g/de q^i/de{q^j}\biggr)_{\!\h\g}\@X^i\@Y^j +
\biggl(\sd\dot g/de q^i/de{z^A}\biggr)_{\!\h\g}\big(X^i\, Y^A+Y^i\,X^A\big)=                    \\[5pt]
\hskip-5pt=\!\biggl[\d C_{ij}/dt+C_{jk}\biggl(\de\psi^k/de{q^i}\biggr)_{\!\h\g}\!\!+C_{ik}\biggl(\de\psi^k/de{q^j}\biggr)_{\!\h\g}\biggr]X^iY^j\!+
C_{ik}\biggl(\de\psi^k/de{z^A}\biggr)_{\!\h\g}\!\Big(X^iY^A\!+X^AY^i\Big).\hskip-5pt
\end{multline}

\smallskip
In particular, if $\@X,\@Y\@$ is any pair of admissible deformations of $\@\g\@$ lifting to deformations $\@\h X,\@\h Y\@$ of $\@\h\g\@$, eqs.~\!(\ref{2.4}),
(\ref{2.5}), together with the variational equation (\ref{1.8}), imply the identity
\begin{equation}\label{2.6}
\d/dt\,\big<\@(\?d^{\,2}g)_{\?\g}\,,\,X\otimes Y\big>\,=\,\d/dt\,\big<\@C\,,\,X\otimes Y\big>\,=\,\big<\@(\?d^{\,2}\dot g\?)_{\?\h\g}\,,\,\h X\otimes\h Y\big>\@.
\end{equation}

\medskip\noindent\vspace{1pt}
\textbf{(\/iii\/)} \,On account of eqs.~\!(\ref{2.2}), (\ref{2.5}), the second derivatives $\@\Big[\sd\Lagr\?'/de z^A\plus70/de{z^B}\?\Big]{\plus05}_{\?\h\g}$
are independent of the choice of $\@\Lagr\?'\@$\vspace{.5pt} within the class of adapted Lagrangians: the action of the functional (\ref{2.3}) on pairs of
vertical vectors $\@\h X= X^A\/\big(\de/de{z^A}\big){\plus05}_{\!\h\g}$, $\@\h Y= Y^A\/\big(\de/de{z^A}\big){\plus05}_{\!\h\g}\@$\vspace{1pt} is therefore
\emph{invariant\/} under gauge transformations of the second kind.

In coordinates, setting $\@\big<\big(d^{\,2}\Lagr\?'\big)_{\@\h\g}\,,\,\h X\otimes\@\h Y\@\big>=G_{AB}\,X^A\@Y^B\@$ and recalling eq.~\!(\ref{2.1}), a direct
calculation yields the expression
\begin{equation}\label{2.7}
G_{AB}:=\bigg[\?\sd(\Lagr-\dot S)/de z^A\plus80/de{z^B}\?\bigg]_{\?\h\g}\!=
\bigg[\sd\Lagr/de z^A/de{z^B}\bigg]_{\!\h\g}\!-\,p_i\/(t)\bigg[\sd\psi^i/de z^A/de{z^B}\bigg]_{\!\h\g}\!:=-\@\bigg[\sd \K/de z^A/de{z^B}\bigg]_{\!\h\g}\@.
\end{equation}
The function
\begin{equation*}
\K\/(t,q^i,z^A):=\,p_i\/(t)\,\psi^i\/(t,q^i,z^A)\@-\@\Lagr\/(t,q^i,z^A)
\end{equation*}
will be called the \emph{restricted Pontryagin Hamiltonian\?}.\vskip1pt

Notice that, on account of the identification $\@\Big[\sd\K/de z^A\plus70/de{z^B}\Big]\/{\plus05}_{\h\g\/(t)}\!=\Big[\sd\Ham/de
z^A\plus70/de{z^B}\Big]\/{\plus05}_{\gt\/(t)}$\vspace{1pt}, the matrix (\ref{2.7}) is non--singular along any \emph{regular\/} extremal. More generally, if
$\@\det G_{AB}\neq 0\@$\vspace{.5pt} on a closed interval $\@[\?a,b\,]\subset [t_0,t_1]\@$, the restriction $\@\g\colon[\?a,b\,]\to\V\@$ will be called a
\emph{regular arc\/} of $\@\g\@$.

\medskip\noindent
\textbf{(\/iv\/)} \,The essential Hessian $\@\big(d^{\,2}\Lagr\?'\?\big)_{\h\g}\@$ determines an infinitesimal control along every regular arc
$\@\g\colon[\?a,b\,]\to\V\@$. The algorithm is not invariant under restricted gauge transformations, but explicitly depends on the choice of the Lagrangian.

To start with we observe that, under the assumption $\@\det G_{AB}\neq 0\@$,\vspace{.5pt} there exists a unique linear section $\@h\colon V\/(\g\/(t))\to
A\/(\h\g\/(t))\@$ satisfying the requirement
\begin{equation}\label{2.8}
\left<\big(\?d^{\,2}\Lagr\?'\?\big)_{\h\g\/(t)}\,,\,h(X)\otimes\h Y\right>\@=\,0\qquad\forall\,X\in
V\/(\g\/(t))\;,\,\h Y\in V\/(\h\g\/(t))\@.
\end{equation}
In coordinates, preserving the notation (\ref{1.9}), eqs.~\!(\ref{2.8}) amount to the set of conditions
\begin{equation*}
\bigg<\big(\?d^{\,2}\Lagr\?'\?\big)_{\h\g}\;,\,\DE_i\otimes\bigg(\de/de{z^A}\@\bigg)_{\!\h\g}\bigg>\@=\,
\bigg(\sd\Lagr\?'/de q^i/de{z^A}\bigg)_{\!\h\g}\!+\,G_{AB}\,\@h\@_i{}^B\,=\,0\@.
\end{equation*}
The latter may be uniquely solved for the coefficients $\@h\@_i{}^B$, yielding the expressions
\begin{equation}\label{2.9}
h\@_i{}^B=-\@G^{BC}\left(\sd\Lagr\@'/de q^i/de{z^C}\right)_{\!\h\g}\@,
\end{equation}
with $\@G_{AB}\,G^{BC}=\delta\@_A^C\@$. The horizontal distribution $\@\Hg=h\?(\Vg)\@$ associated with $\@h\@$ is therefore spanned by the vector
fields\vspace{2pt}
\begin{equation*}
\DE_i\,:=\,h\left(\de/de{q^i}\right)_{\!\g}=\,\bigg(\de/de{q^i}\bigg)_{\!\h\g}-
 G^{AB}\bigg(\sd\Lagr\@'/de q^i/de{z^B}\bigg)_{\!\h\g}\bigg(\de/de{z^A}\bigg)_{\!\h\g}\@.
\end{equation*}
The vectors $\@\DE_i\@,\@\big(\de/de{z^A}\big)_{\!\h\g}\@$ provide a basis for $\@\A\/(\h\g\/(t))\@$ at each $\@t\in[\?a,b\,]\@$.
In terms of this basis, setting $\@\h X=X^i\@\DE_i+U^A\big(\de/de{z^A}\big)_{\h\g\/(t)}\@,\,\h Y= Y^i\@\DE_i+V^A\big(\de/de{z^A}\big)_{\h\g\/(t)}\@$, the
representation (\ref{2.3}) of the essential Hessian simplifies to
\begin{equation}\label{2.10}
\Big<\big(d^{\,2}\Lagr\?'\big)_{\@\h\g\/(t)}\,,\,\h X\otimes \h Y\@\Big>\,=\,N_{ij}\,X^i\@Y^j+\,G_{AB}\,U^A\@V^B\@,\vspace{3pt}
\end{equation}
with
\begin{equation}\label{2.11}
U^A=X^A+\@G^{AB}\biggl(\sd\Lagr\@'/de q^i/de{z^B}\biggr)_{\!\h\g}\@X^i,\quad\;V^A=Y^A+\@G^{AB}\biggl(\sd\Lagr\@'/de q^i/de{z^B}\biggr)_{\!\h\g}\@Y^i
\end{equation}
and
\begin{equation}\label{2.12}
N_{ij}=\bigg(\sd\Lagr\?'/de q^i/de{q^j}\bigg)_{\!\h\g}\!-\,G^{AB}
\bigg(\sd\Lagr\?'/de q^i/de{z^A}\bigg)_{\!\h\g}\!\bigg(\sd\Lagr\?'/de q^j/de{z^B}\bigg)_{\!\h\g}
\end{equation}

\vskip2pt
The absolute time derivative along $\@\g\/\big([\?a,b\,])\@$ induced by $\@h\@$ will be denoted by $\!\D/D{\plus70t}\@$\vspace{1pt}.
\linebreak
The expression (\ref{1.11}) for the temporal connection coefficients takes now the form
\begin{equation}\label{2.13}
\tau\?_k{}^i\,:=\,-\,\DE_k\!\left(\psi^i\right)=\,-\bigg(\de\@\psi^i/de{q^k}\bigg)_{\!\h\g}
+\, G^{AB}\bigg(\de\@\psi^i/de{z^A}\bigg)_{\!\h\g}\bigg(\sd\Lagr\@'/de q^k/de{z^B}\bigg)_{\!\h\g}\@.
\end{equation}

\medskip\noindent
\textbf{(\/v\/)} \,The coefficients (\ref{2.12}) form the components of a symmetric virtual tensor\linebreak $\@N=N_{ij}\,\wg{i}\!\otimes\wg{j}\@$ along
$\@\g\/\big([\?a,b\,])\@$, uniquely defined by the prescription
\begin{equation*}
\Big<N\?,\@X\otimes Y\Big>\@=\@\Big<(d^{\,2}\Lagr\?')_{\@\h\g}\@,\@h\/(X)\otimes h\/(Y)\Big>\qquad\forall\;X,Y\in V\/(\g\/(t))\@.
\end{equation*}

Under a gauge transformation of the second kind (\ref{2.2}), eqs.~\!(\ref{2.5}), (\ref{2.12}) entail the transformation law
\begin{multline*}
N_{ij}\@\to\@\ovl N_{ij}\@=\@N_{ij}\@-\,\d C_{ij}/dt\@-\@C_{jk}\biggl(\de\psi^k/de{q^i}\biggr)_{\!\h\g\/(t)}\!-\@
C_{ik}\biggl(\de\psi^k/de{q^j}\biggr)_{\!\h\g\/(t)}\!+                                                                                      \\[2pt]
+\,G^{AB}\bigg[C_{jk}\bigg(\sd\Lagr\@'/de q^i/de{z^A}\,\de\@\psi^k/de{z^B}\biggr)_{\!\h\g\/(t)}\!+\@ C_{ik}\bigg(\sd\Lagr\?'/de
q^j/de{z^B}\,\de\@\psi^k/de{z^A}\biggr)_{\!\h\g\/(t)}\!-\biggl(\de\@\psi^h/de{z^A}\,
\de\@\psi^k/de{z^B}\biggr)_{\!\h\g\/(t)}C_{ih}\,C_{jk}\bigg]\@.\vspace{1pt}
\end{multline*}

\smallskip\noindent
Setting
\begin{equation*}
M^{rs}\@:=\@G^{AB}\bigg(\de\psi^r/de{z^A}\bigg)_{\!\h\g}\bigg(\de\psi^s/de{z^B}\bigg)_{\!\h\g}
\end{equation*}
and recalling eqs.~\!(\ref{1.12}), (\ref{2.13}) as well as the symmetry of $\@C_{ij}\@$, the latter may be synthetically written as
\begin{equation}\label{2.14}
\ovl N_{ij}\@=\@N_{ij}-\D\?C_{ij}/Dt-\@ M^{rs}\@C_{ir}\,C_{sj}\@,
\end{equation}
$\!\D/D{\plus70t}\@$\vspace{1pt} denoting the absolute time derivative along $\@\g\@$ induced by the infinitesimal control associated with $\@\Lagr\?'$. Hence
the result:
\begin{proposition}\label{Pro2.1}
Let $\@\g\colon[\?a,b\,]\to\V\@$ be a regular arc of a normal extremal. Then, through a suitable gauge transformation of the second kind, the essential Hessian
(\ref{2.10}) may be reduced to the canonical form
\begin{equation}\label{2.15}
\Big<\@\big[\?d^{\,2}(\Lagr\?'-\dot g)\?\big]_{\h\g}\,,\,\h X\otimes\h Y\@\Big>\,=\,
G_{AB}\,\ovl U^A\,\ovl V^B
\end{equation}
in a neighborhood of each point $\@t^*\in [\?a,b\,]$, with the components $\@\ovl U^A,\ovl V^A\@$\vspace{1pt} related to the components (\ref{2.11}) by the
linear transformation
\begin{equation*}
\ovl U^A=U^A\!-\@G^{AB}\Big(\sd\dot g/de q^i/de{z^B}\Big)_{\!\h\g}\@X^i,\quad\qquad\ovl V^A=
V^A-\@G^{AB}\Big(\sd\dot g/de q^i/de{z^B}\Big)_{\!\h\g}\@Y^i\@.
\end{equation*}
\end{proposition}
\begin{proof}
The conclusion follows at once from eq.~\!(\ref{2.14}) observing that the (symmetric) matrix equation $\@\ovl N_{ij}\@=\@0\@$ is always solvable for the
unknown $\@C_{ij}\/(t)\@$ in a neighborhood of $\@t^*\@$.
\end{proof}
\smallskip
\subsection{The Legendre condition}\label{S2.2}
We now apply the previous results to the study of the second variation of the action functional along a locally normal extremal $\@\g\@$. To~this~end,~we
replace the original Lagrangian $\@\Lagr\@$ with a gauge equivalent one, arbitrarily chosen within the class of adapted Lagrangians.
As already pointed out, the soundness of the procedure relies on the fact that, for any $\@S\in\F\/(\V)\@$ and for each admissible deformation $\@\g_\xi\@$
with fixed endpoints, the functions $\@\I\?[\g_\xi]= \int_{\h\g_\xi}\Lagr\@\@dt\@$ and $\@\I\?'[\g_\xi]=\int_{\h\g_\xi}(\Lagr-\dot S)\@\@dt\@$ differ by a
constant, and have therefore the same second derivatives.

In particular, when $\@\Lagr\?'\@$ is adapted to $\@\g\@$, the conditions $\@\big(\de\Lagr\?'/de{q^k}\big)_{\!\h\g}\@
=\big(\de\Lagr\?'/de{z^A}\big)_{\!\h\g}\@=\@0\@$, together with eq.~\!(\ref{2.3}), yield the plainly covariant result
\begin{multline}\label{2.16}
\frac{d^{\,2}\I\?[\g_{\xi}]}{d \xi^2}\bigg|_{\xi=0}\,=\,\int_{t_0}^{t_1}\bigg[\bigg(\sd\Lagr\?'/de
q^i/de{q^j}\bigg)_{\!\h\g}\,X^i\,X^j\,+\,2\,\bigg(\sd\Lagr\?'/de q^i/de{z^A}\bigg)_{\!\h\g}\,X^i\,X^A\,+  \\[3pt]
+\,\bigg(\sd\Lagr\?'/de z^A/de{z^B}\bigg)_{\!\h\g}\,X^A\,X^B\bigg]\,d\/t\,=\,
\int_{t_0}^{t_1}\Big<\big(d^{\,2}\Lagr\?'\big)_{\@\h\g}\,,\,\h X\otimes\h X\@\Big>\,d\/t\@, \hskip.4cm
\end{multline}
$\h X=X^i\big(\de/de{q^i}\big)_{\!\h\g}\!+X^A\big(\de/de{z^A}\big)_{\!\h\g}$ denoting the infinitesimal deformation associated with~$\h\g_\xi$.
Under gauge transformations of the second kind, eq.~\!(\ref{2.6}) entails the transformation law
\begin{equation*}
\big<\@\big[\?d^{\,2}(\Lagr\?'-\dot g\?)\?\big]_{\@\h\g}\,,\h X\otimes\h X\@\big>=
\big<\big(\?d^{\,2}\Lagr\?'\?\big)_{\@\h\g}\,,\h X\otimes\h X\@\big>-
\d/dt\,\big<\?(\?d^{\,2}g\?)_{\@\g}\,,X\otimes X\big>\@,
\end{equation*}
confirming the gauge invariance of the integral (\ref{2.16}) within the class of fixed endpoints deformations, but pointing out the non--invariance of the
integrand.

Proposition \ref{Pro2.1} plays a role in the identification of a necessary condition for the extremal $\@\g\@$ to yield a minimum for the action functional.
A useful result in this sense is provided by the following
\begin{lemma}\label{Lem2.1}
Given a normal extremal $\g\colon[\@t_0\@,t_1]\to\V$, take any vertical vector\linebreak $V=V^A\@\big(\de/de{z^A}\big)_{\h\g\/(t^*)}$\vspace{1pt} at a point
$\h\g\/(t^*)$, $t^*\!\in(\@t_0\@,t_1)$. Then, if $\@G_{AB}\/(t^*)\,V^A\?V^B\neq 0$, there exists an infinitesimal deformation $X$\vspace{1pt} of $\g$ vanishing at the
endpoints such that the second variation $\@\int_{t_0}^{t_1}\big<\big(d^{\,2}\Lagr\?'\big)_{\@\h\g}\,,\,\h X\otimes\h X\@\big>\,d\/t\@$ has the same sign as
$\@G_{AB}\/(t^*)\,V^A\@V^B$.
\end{lemma}
\begin{proof}
We extend $V$ to a vector field with compact support along $\h\g$ and choose $\eps' >0$ small enough as to ensure $G_{AB}\@V^A\@V^B\neq0\@$ for all
$|\@t-t^*|<\eps'\@$.

Depending on the value of $\det G_{AB}\/(t^*)$ we consider the following two cases:

\noindent
\textit{i)}\, if $\det G_{AB}\/(t^*) \neq 0\@$, on account of Proposition \ref{Pro2.1}, there exist $\eps\le\eps'$ and a gauge transformation of the second
kind $\@\Lagr\?''=\Lagr\?'-\dot g\@$ such that both the condition $\det G_{AB}\/(t)\neq 0\@$ and the representation (\ref{2.15}) hold throughout the interval
$\@t^*\!-\eps<t<t^*\!+\eps\@$.

Given any $[\/c\@,d\,] \subset (\@t^*\!-\eps\@,t^*\!+\eps)\@$, denote by $\g'$ the regular arc $\@\g\colon[\/c\@,d\,]\to\V\@$, by $\@\ovl{h}\colon V\/(\g')\to
A(\h\g')\@$ the infinitesimal control along $\g'$ determined by the essential Hessian $\@\big(d^{\,2}\Lagr\?''\?\big)_{\h\g}\@$, and by $\ovl{\plus50e}_\na\@$
a corresponding $\@\ovl{h}\@$--transported basis.

Also, let $\vphi\/(t)\@$ denote a differentiable function with compact support contained in $[\/c\@,d\,]\@$, satisfying the properties\@
\footnote%
{\@The existence of such a $\@\vphi\/(t)\@$ follows from elementary distribution theory.}:
\begin{equation*}
\vphi\/(t^*)=1\quad,\qquad
\int_c^d\ovl{\plus50e}^\na{}_i\@\bigg(\de\psi^i/de{z^A}\bigg)_{\h\g}\vphi\ V^A \,d\/t = 0\@.
\end{equation*}
Setting
\begin{equation*}
\ovl{U}^A\/(t)\ :=\
\begin{cases}
\vphi\/(t)\,V^A\/(t) & t \in [\/c\@,d\,]                                                            \\[3pt]
0 & t \notin [\/c\@,d\,]
\end{cases}\ ,
\quad X^\na\/(t)\ :=\ \int_{t_0}^t \ovl{\plus50e}^\na{}_i\@
\bigg(\de \psi^i /de {z^A}\bigg)_{\h\g\/(t)} \ovl{U}^A \,d\/t
\end{equation*}
it is readily seen that the field $X = X^\na\/(t)\,\ovl{\plus50e}_\na\@$ is an infinitesimal deformation of $\g$ with support contained in $[\/c\@,d\,]$,
lifting to $ \h X = X^\na\,\ovl{h}\@(\/\ovl{\plus50e}_\na\/) + \ovl{U}^A\big(\de /de {z^A}\big)_{\h\g}\@$.

Collecting all results and recalling the gauge invariance of the integral (\ref{2.16}), we conclude that the expression
\begin{equation*}
\int_{t_0}^{t_1}\Big<\big(d^{\,2}\Lagr\?'\big)_{\@\h\g}\,,\,\h X\otimes\h X\@\Big>\,d\/t\, =\,
\int_c^d G_{AB}\, \ovl{U}^A\@ \ovl{U}^B\,d\/t\, = \int_c^d \vphi^2\@ G_{AB}\, V^A\@ V^B\,d\/t
\end{equation*}
has the same sign as $G_{AB}\/(t^*)\, V^A\@V^B\@$.

\medskip\noindent
\textit{ii)}\, if $\det G_{AB}\/(t^*) = 0\@$, we introduce the auxiliary Lagrangian
\begin{equation*}
\@\mathscr{M}\,=\,\Lagr\?'\,-\,
\l\,\delta_{AB}\@\Big(z^A-z^A\/\big(\h\g(t)\big)\Big)\@\Big(z^B-z^B\/\big(\h\g(t)\big)\Big)\@,
\end{equation*}
with  $\@\l\in\R$ chosen in such a way as to ensure the validity of the conditions
\begin{equation*}
\det\big(G_{AB}\/(t^*)-\l\,\delta_{AB}\big)\neq 0\;,\quad \frac{G_{AB}\/(t^*)\,V^A\@V^B}{\l}\,>\,\delta_{AB}\,V^A\@V^B.
\end{equation*}

A straightforward check shows that $\@\g$ is an extremal for the action functional $\@\int_{\h\g} \mathscr{M}\,d\/t\@$, that $\@\mathscr{M}$ is adapted to
$\@\g$ and that, for each $\h X = X^i \big(\de /de {q^i}\big)_{\h\g} + X^A \big( \de /de {z^A}\big)_{\h\g}\@$, the essential Hessian of $\mathscr{M}$ satisfies
the relation
\begin{equation}\label{2.17}
\Big<\big(d^{\,2}\mathscr{M}\big)_{\@\h\g}\,,\,\h X\otimes\h X\@\Big>\,=\,
\Big<\big(d^{\,2}\Lagr\?'\big)_{\@\h\g}\,,\,\h X\otimes\h X\@\Big>\,-\,\l\,\delta_{AB}\ X^A\@X^B.
\end{equation}

\noindent
Setting $G'_{AB}\/(t):=\big(\sd\mathscr{M}/de{z^A}/de{z^B}\big)_{\h\g}=\@G_{AB}\/(t)-\l\@\delta_{AB}\@$\vspace{1pt} we have then the properties\?:\vspace{4pt}
\\
$\bullet\;$ the matrix $G'_{AB}\/(t^*)$ is non--singular;\vspace{2pt}
\\
$\bullet\;$ both expressions $G_{AB}\/(t^*)\,V^A\@V^B\@$ and $G'_{AB}\/(t^*)\, V^A\@V^B\@$ have the same sign as $\l\@$.\vspace{3pt}

According to our previous discussion, we can therefore find an infinitesimal deformation $X$ vanishing at the endpoints, such that
$\@\int_{t_0}^{t_1}\big<\big(d^{\,2}\mathscr{M}\big)_{\@\h\g}\,,\,\h X\otimes\h X\@\big>\, d\/t\@$ has the same sign as $\@G'_{AB}\/(t^*)\,V^A\@V^B\@$.
On account of eq.~\!(\ref{2.17}), the expression
\begin{equation*}
\int_{t_0}^{t_1}\!\Big<\big(d^{\,2}\Lagr\?'\big)_{\@\h\g}\,,\,\h X\otimes\h X\@\Big>\,d\/t\,=\,
\int_{t_0}^{t_1}\!\Big<\big(d^{\,2}\mathscr{M}\big)_{\@\h\g}\,,\,\h X\otimes\h X\@\Big>\,d\/t\,+\,\l\@\int_{t_0}^{t_1}\delta_{AB}\,X^A\@X^B\,d\/t
\end{equation*}
has the same sign as $\@G_{AB}\/(t^*)\,V^A\@V^B$.
\end{proof}

\noindent
As an immediate consequence of Lemma \ref{Lem2.1} we have
\begin{corollary}[\/\bf{Legendre condition}\/]\label{Cor2.1}
A necessary condition for a normal extremal\linebreak $\@\g\colon[\@t_0,t_1]\to\V\@$ to yield a local minimum for the action functional is the positive semi-definiteness
of the matrix $\@G_{AB}\/(t)\@$ at all $\@t\in[\@t_0,t_1]\@$.
\end{corollary}
\subsection{Regular extremals}\label{S2.3}
An especially remarkable situation occurs when $\@\g\@$ is a \emph{regular\/} extremal, namely a normal extremal satisfying $\det G_{AB}\/(t)\ne 0\;\,\forall\;
t\in[\/t_0\@,\@t_1\/]\@$. Corollary \!\ref{Cor2.1} then specializes into the following
\begin{corollary}[\/\bf{Strengthened Legendre condition}\/]\label{Cor2.2}
A necessary condition for a regular normal extremal $\@\g\colon[\@t_0,t_1]\to\V\@$ to yield a minimum for the action functional is that the matrix
$\@G_{AB}\/(t)\@$ be positive definite at all $\@t\in[\@t_0,t_1]\@$.
\end{corollary}

All result concerning regular arcs established in Sec.\!~\ref{S2.1} now apply to the whole of $\g\@$. In particular, according to Proposition \ref{Pro2.1}, for
each $\@t^*\in[\?t_0,t_1\?]\@$ there exists a gauge transformation of the second kind $\@\Lagr\?'\to\Lagr\?'-\dot g\@$ satisfying
\begin{equation}\label{2.18}
\Big<\@\big[\?d^{\,2}(\Lagr\?'-\dot g)\?\big]_{\h\g\/(t)}\,,\,\h X\otimes\h X\@\Big>\,=\,G_{AB}\/(t)\,\ovl U^A\,\ovl U^B\qquad\forall\;t\in [\?a,b\?]\@.
\end{equation}
for all $\@t\@$ in a neighborhood of $\@t^*$.

Unfortunately, the purely local character of this result is of little help in the study of the second variation (\ref{2.16}). An important issue is therefore
establishing under what circumstances eq.~\!(\ref{2.18}) holds over the whole interval $\@[\@t_0,t_1]\@$.
On account of eqs.~\!(\ref{2.10}), (\ref{2.14}), this means analysing the interval of existence of the solutions of the Riccati--like differential equation
\begin{equation}\label{2.19}
\D\?C_{ij}/Dt\,+\,M^{rs}\,C_{ir}\,C_{sj}\,-\,N_{ij}\,=\,0
\end{equation}
for the unknown $\@C_{ij}\/(t)\@$.

\smallskip
A significant insight into this problem is provided by the following

\begin{theorem}\label{Teo2.1}
Let $\g$ be a locally normal extremal, carrying a positive definite matrix $\@G_{AB}\/(t)\/$, $t\in[\@t_0,t_1]$.
Let $K = K^i{}_j\/(t)\big(\de /de {q^i}\big)_\g\!\otimes\@\wg{j}$, $E = E_{ij}\/(t)\,\wg{i}\otimes\wg{j}$ be two virtual tensors along $\g\@$ obeying the
transport laws
\begin{equation}\label{2.20}
\D K^i{}_j/Dt\,=\,M^{ir}\,E_{rj}\,,\qquad\quad \D E_{ij}/Dt\, =\,N_{ir}\,K^r{}_j
\end{equation}
with initial data satisfying the conditions $K^i{}_j\/(t_0)=0\@$, $\det E_{ij}\/(t_0) \neq 0\@$.

\noindent
For any $a \in (\/t_0\@,\/t_1\/]\@$, the following statements are then equivalent\@:

\begin{itemize}
\item[\textit{(i)}]  eq.~\!(\ref{2.19}) admits a regular solution throughout the interval $[\/t_0\@,\/a\/]\@$;
\smallskip
\item[\textit{(ii)}] \vspace{1pt}$\det K^i{}_j\/(t) \neq 0\ \forall\ t \in (\/t_0\@,\/a\/]\@$.
\end{itemize}
\end{theorem}

\begin{proof}
To avoid ambiguities, for each $\tau\in(\/t_0\@,\/a\/]\@$ we denote by $\g_\tau$ the closed arc $\g\colon[\/t_0\@,\/\tau\/]\to\V\@$. Due to the stated
assumptions, $\g_\tau\@$ is then a normal extremal of the action functional. Bearing this in mind, let us now come to the core of the proof.

\smallskip \noindent
\textit{(i)\,$\Rightarrow$(ii)}\ \, For any $\tau \in (\/t_0\@,\/a\/]\@$, the stated hypotheses entail
\begin{equation}\label{2.21}
\int_{t_0}^\tau\,\Big<\big(d^{\,2}\Lagr\?'\big)_{\@{\h\g}_\tau}\,,\,\h X\otimes\h X\@\Big>\,dt\ >\ 0
\end{equation}
for any non--null section $\hat X \colon [\/t_0\@,\/\tau\/]\to\A({\h\g}_\tau)\@$ arising from the lift of a corresponding infinitesimal variation $X$ of
$\@\g_\tau\@$ vanishing at the endpoints.

\noindent
We claim that, as a consequence of this fact, the linear map $\@V\/(\g\/(\tau))\to V\/(\g\/(\tau))\@$ determined by the tensor $\@K\/(\tau)\@$ is necessarily
injective.

To verify this assertion, given any vector $\@\b=\b\@^j\@\big(\de/de{q^j}\big)_{\g\/(\tau)}\in\Ker(K\/(\tau))\@$\vspace{-.4pt}, we prolong it to a
$\@h\@$-transported section $\@\b\colon[\/t_0\@,\/\tau\/]\to V(\g_\tau)\@$\vspace{-.4pt}, i.e.~to a vector field $\@\b\@^j\/(t)\@\big(\de/de{q^j}\big)_\g\@$
satisfying $\@\D{\b^j}/Dt=0\@$. On account of eqs.~\!(\ref{2.20}) it is then readily seen that the fields\vspace{0pt}
\begin{equation*}
\quad X:=K^i{}_j\/(t)\,\b\@^j\bigg(\de/de {q^i}\bigg)_{\!\g_\tau}\!\!:=X^i\/(t)\@\bigg(\de/de{q^i}\bigg)_{\!\g_\tau}\!,
\qquad \l:=E_{ij}\/(t)\,\b\@^j\,\wg{i}:= \l_{\@i}\/(t)\,\wg{i}
\end{equation*}
fulfil the evolution equations
\begin{subequations}\label{2.22}
\begin{align}
\D X^i/Dt\,&=\,M^{ir}\,E_{rj}\,\b\@^j\@=\@M^{ir}\,\l_{\@r}\,,                               \\[5pt]
\D \@\l_{\@i}/Dt\, &=\,N_{ir}\,K^r{}_j\,\b\@^j\@=\@N_{ir}\,X^r.\hskip1.5cm
\end{align}
\end{subequations}

Let us now prove that the \emph{unique\/} solution of eqs.~\!(\ref{2.22}\@a,\@b) consistent with the requirements $\@X^i\/(t_0)=X^i\/(\tau)=0\@$ is the null
one. To this end we set
\begin{equation}\label{2.23}
U^A:=\,G^{AB}\biggl(\de\psi^r/de{z^B}\biggr)_{\!\h\g_\tau}\/\l_{\@r}
\end{equation}
and observe that, with this definition, the vector field $\@\h X:=X^i\@\DE_i\@ +\@U^A\big(\de/de{z^A}\big)_{\!\h\g_\tau}\@$ satisfies the variational equation
\begin{equation*}
\D X^i/Dt\,=\@M^{ir}\,\l_{\@r}\,=\@G^{AB}\,\de\psi^i/de{z^A}\,\de\psi^r/de{z^B}\;\l_{\@r}\,=\,\de\psi^i/de{z^A}\;U^A
\end{equation*}
as well as the identity
\begin{equation*}
\Big<\big(d^{\,2}\Lagr\?'\big)_{\@{\h\g}_\tau}\,,\,\h X\otimes\h X\@\Big>\@=\@
N_{ij}\,X^i\@X^j+\,G_{AB}\,U^A\@U^B=\@\D \@\l_{\@i}/D t\@X^i\@+\,\l_{\@i}\@\D X^i/D t\@=\@\d/dt\@(\/\l_{\@i}\@X^i\/)\@.
\end{equation*}

Therefore, $\@\h X\@$ is the lift of an admissible infinitesimal deformation $X$ of $\@\g_\tau$, vanishing at the endpoints and satisfying
\begin{equation*}
\int_{t_0}^\tau\, \Big<\big(d^{\,2}\Lagr\?'\big)_{\@{\h\g}_\tau}\@,\@\h X\otimes\h X\@\Big>\,dt\ =\ \l_{\@i}\@X^i\@\big\lvert^\tau_{t_0}\,=\, 0\,.
\end{equation*}

On account of eq.~\!(\ref{2.21}), this entails $\@\h X\/(t)=0\;\Longleftrightarrow\;X^i\/(t)=U^A\/(t)=0\@$. Eqs.~(\ref{2.22}\@b), (\ref{2.23}) take then the
form
\begin{equation*}
\D\l_{\@i}/Dt\,=\,0\,,\qquad\; \l_{\@r}\bigg(\de\psi^r/de{z^A}\bigg)_{\!\h\g}=\,0\@,
\end{equation*}
mathematically equivalent to  $\@\l_{\@i}\/(t)=0\@$ because of the local normality of $\g\@$, as expressed in Remark \ref{Rem1.2}\/.

Collecting all results and recalling the relation $\@\l_{\@i}=E_{ij}\,\b\@^j\@$ as well as the assumption $\@\det E_{ij}\/(t_0)\ne 0\@$ we conclude that
$\@\b\@^j\/(t_0)=0\@$.

By the very definition of $\@h\@$--transport, this entails $\@\b\@^j\/(t)=0\;\forall\,\@t\@$, whence, in particular, $\@\b\@^j\/(\tau)=0\@$. Therefore,
$\Ker(K\/(\tau))=\{\?0\?\}\@$ $\,\Longrightarrow\;\det K^i{}_j\/(\tau)\ne 0\@$.
By the arbitrariness of $\@\tau\@$, this completes the proof.

\medskip \noindent
\textit{(ii)\,$\Rightarrow$(i)}\ \, On grounds of continuity, property (\/ii) implies $\det K^i{}_j\/(t)\neq0\@$ on a broader interval $(\/t_0\@,\/b\@]\@$,
$b>a$. Straightforward consequences of this fact are:
\begin{subequations}\label{2.24}
\begin{itemize}
\item
the unique solution of eqs.~\!(\ref{2.22}) satisfying $\@X^i\/(t_0)=X^i\/(b)=0\@$ is the null one\@\vspace{2pt}
\footnote%
{\@Due to non vanishing of $\@\det E_{ij}\/(t_0)\ne 0\@$, the most general solution of eqs.~\!(\ref{2.22}) satisfying $\@X^i\/(t_0)=0\@$ is in fact necessarily
of the form $\@X^i\/(t)=K^i{}_j\@\b\@^j\@$, with $\@\D{\b^j}/Dt=0\@$.};
\item
the tensor $C_{ij} := E_{ip}\@(\/K^{-1}\/)^p{}_j\@$ is well-defined on $(\/t_0\@,\/b\@]\@$ and fulfils the equation
\begin{multline}
\D C_{ij}/D t\@=\@\D E_{ip}/D t\@(\/K^{-1}\/)^p{}_j\@+\@E_{ip}\@\D\@(\/K^{-1}\/)^p{}_j /D t\@=         \\
=\@N_{ij}-E_{ip}\@(\/K^{-1}\/)^p{}_r\@M^{rl}\@E_{ls}\@(\/K^{-1}\/)^s{}_j\@=\@ N_{ij}\@-\@C_{ir}\@M^{rl}\@C_{lj}\@,\quad
\end{multline}
formally identical to eq.~\!(\ref{2.19});
\item
the tensor $C_{ij}\@$ is symmetric: its inverse $B^{ij}:=K^i{}_r\@(\/E^{-1}\/)^{rj}\@$ is in fact well--defined in a neighborhood of $t_0\@$, where it fulfils
the transport law
\begin{multline}
\D B^{ij} /D t\@=\@ \D K^i{}_r /D t\@ (\/E^{-1}\/)^{rj}\@+\@K^i{}_r\@ \D \@(\/E^{-1}\/)^{rj} /D t\@=    \\
=\@M^{ij}-K^i{}_r\@(\/E^{-1}\/)^{rq}\@N_{qp}\@K^p{}_s\@(\/E^{-1}\/)^{sj}\@=\@ M^{ij}\@-\@B^{iq}\@N_{qp}\@B^{pj}\@.\;
\end{multline}
The latter, along with the initial data $B^{ij}\/(t_0)=0\@$ and the symmetric character of both matrices $M^{ij}\@$ and $N_{ij}\@$, entails the symmetry of
$B^{ij}\/(t)\@$. This, in turn, ensures the symmetry of $C_{ij}\/(t)\@$ for $\@t\@$ close to $t_0$, and therefore also the symmetry of $C_{ij}\/(t\@)\@$ all
along its definition interval.
\end{itemize}
\end{subequations}

\noindent
To sum up, we proved that, as a consequence of assumption (ii), eq.~\!(\ref{2.19}) admits a symmetric regular solution $C_{ij}\/(t)$ in any closed interval
$[\/c\@,\/b\@]\subset (\/t_0\@,\/b\@]\@$.

However, we are not done yet since, due to the request $K^i{}_j\/(t_0)=0\@$, the tensor $C_{ij}\/(t)\@$ is singular at $t = t_0\@$. In order to overcome this
aspect, we now introduce two solutions $\bar{K}^i{}_j\/(t)\@$, $\bar{E}_{ij}\/(t)\@$ of the (time--reversed) system (\ref{2.20}), subject to the conditions
$\bar{K}^i{}_j\/(b) = 0\@$, $\det \bar{E}_{ij}\/(b)\neq0\@$.

As the Riccati--like matrix equation (\ref{2.19}) admits a regular solution throughout the interval $[\/c\@,\/b\@]$, by virtue of the (already proved)
implication (i)$\@\Rightarrow$(ii) we conclude that\linebreak $\det\bar{K}^i{}_j\/(t)\neq 0\;\forall\,t\in[\/c\@,\/b\@)\@$. By the arbitrariness of $\@c\@$ and by the
request $\@a<b\@$ this entails $\@\det\bar{K}^i{}_j\/(t)\neq 0$ $\@\forall\,t\in(\/t_0\@,\/a\@]\@$.

Let us now prove that $\det\bar K^i{}_j\/(t_0)\@$ cannot vanish. To this end, given any solution $\@\bar\b\@^1\And\bar\b\@^n\@$ of the linear homogeneous
system $\bar{K}^i{}_j\/(t_0)\,\bar\b\@^j=0\@$, we prolong it to a $\@h\@$-transported vector field $\@\bar\b\@^j\/(t)\@\big(\de /de {q^j}\big)_\g\@$ along
$\@\g\@$.
\\[1pt]
The fields $\@X^i=\bar K^i{}_j\/(t)\,\bar\b\@^j\/(t)$, $\@\l_{\@i}=\bar E_{ij}\/(t)\,\bar\b\@^j\/(t)\@$ are then readily seen to fulfil the transport law
(\ref{2.22}\@a,\@b) as well as the conditions $\@X^i\/(t_0)=X^i\/(b)=0\@$.
But, as already pointed out, this implies $\@X^i\/(t)=\l_{\@i}\/(t)=0\@$, whence also $\@\bar\b\@^i=0\@$, thereby establishing the non singularity of $\@\bar
K^i{}_j\/(t_0)\@$.

Collecting all results and arguing as before we conclude that, whenever property (ii) holds, the tensor $\bar C_{ij}=\bar E_{ip}\@(\/\bar K^{-1}\/)^p{}_j\@$ is
well-defined all over the interval $[\/t_0\@,\/b\@)$ and thus also over $[\/t_0\@,\/a\/]\@$, is symmetric, and fulfils the transport law
eq.~\!(\ref{2.24}\@a)\@, formally identical to the Riccati--like equation (\ref{2.19})\@
\footnote%
{\@The idea of relating the solutions of the non--linear matrix-Riccati equation \eqref{2.19} to those of the coupled linear systems \eqref{2.20} goes back to
Radon (see \cite{Radon1, Radon2}) and to Reid (see \cite{Reid1}).}.
\end{proof}

The content of Theorem \ref{Teo2.1} is enhanced and made more transparent by introducing the concept of \emph{conjugate point}.

\begin{definition}\label{Def2.1}
Given a locally normal, regular extremal $\@\g\colon[\@t_0\@,t_1]\to\V\@$, let $K = K^i{}_j\/(t)\@\big(\de/de {q^i}\big)_\g\/\otimes\,\wg{j}\,$,
$E=E_{ij}\/(t)\,\wg{\@i}\@\otimes\@\wg{\@j}\@$ be two virtual tensors along $\g\@$, obeying the transport laws (\ref{2.20}) with initial data satisfying the
conditions $K^i{}_j\/(t_0)=0\@$, $\det E_{ij}\/(t_0) \neq 0\@$.\vspace{1pt} A point $\@\tau\in(\@t_0,t_1]\@$ is then said to be \emph{conjugate} to $\@t_0\@$
along $\@\g\@$ if and only if\?:
\begin{itemize}
\item
$\@\det\/K^i{}_j\/(t)\neq0\@$ $\@\forall\,\@t\in(\@t_0\,,\@\tau\@)\@$;
\vskip1pt
\item
$\det\/K^i{}_j\/(\tau) = 0$.
\end{itemize}
\end{definition}
The soundness of Definition \ref{Def2.1} is ensured by the fact that the zeroes of $\det\/K^i{}_j\/(t)\@$ are independent of the specific choice of the initial
values $\@E_{ij}\/(t_0)\/$.
Changing $\@E_{ij}\/(t_0)\@$ into $\@E_{ir}\/(t_0)\@A^r{}_j\@$, with $\@\det\/A^r{}_j\ne0\@$, is in fact reflected into a transformation $\@K^i{}_j\/(t)\to
K^i{}_r\/(t)\@\@A^r{}_j\@$, $\@E_{ij}\/(t)\to E_{ir}\/(t)\@A^r{}_j\@$ of the resulting fields.

With the terminology of Definition \ref{Def2.1}, Theorem \ref{Teo2.1} entails the following
\begin{corollary}\label{Cor2.3}
Let $\g\colon [\@t_0 , t_1] \to \V\@$ be a locally normal extremal, carrying a positive definite matrix $\@G_{AB}\/(t)\@$. Then:
\begin{itemize}
\item[\textit{(i)}] 
a necessary condition for $\g$ to represent a minimum of the action functional is the absence of conjugate points to $\@t_0\@$ throughout the open interval
$(\@t_0,t_1\?)$;
\item[\textit{(ii)}] \vspace{2pt}
a sufficient condition for $\g$ to represent a minimum of the same functional is the absence of conjugate points to $\@t_0\@$ throughout the half--closed
interval $(\@t_0,t_1\?]\@$.
\end{itemize}
\end{corollary}
\begin{proof}
Assertion (i) is established by contradiction. The argument is an adaptation of a classical result by Bliss, as presented e.g.~in \cite{Sagan}\@.

From the proof of Theorem \ref{Teo2.1} we know that if a point $\tau\in(\@t_0,t_1\?)\@$ fulfils the conjugacy condition stated in Definition \ref{Def2.1},
eqs.~(\ref{2.22}\@a,\@b) admit a solution $\@X^i,\l_{\@i}\@$ satisfying\linebreak
$\@X^i\/(t_0)=X^i\/(\tau)=0\@$, $\@X^i\/(t)\ne 0\@$
$\,\forall\,\@t\in(\@t_0,\tau\?)\@$. In particular:
\begin{itemize}
\item
the field $X$ is an admissible infinitesimal deformation of $\g$, vanishing at $t_0$ and at $\tau$ (\@but not necessarily at $t_1$\/), whose lift is given by
$\@\h X:=X^i\@\DE_i\@+\@U^A\big(\de/de{z^A}\big)_{\!\h\g}\@$, with $U^A\/(t)=G^{AB}\big(\de{\psi^r}/de{z^B}\big)_{\h\g}\@\l_{\@r}\/(t)\@$;\vspace{1pt}
\item
given any infinitesimal deformation $Y\@$ vanishing at the endpoints of $\g$ and denoted by $\h Y=Y^i\@\DE_i\@ +\@V^A\big(\de/de{z^A}\big)_{\!\h\g}\@$ the
corresponding lift, eqs.~\!(\ref{2.10}), (\ref{2.22}\@b), (\ref{2.23}) and the variational equation (\ref{1.13}\@a) for the field $\@\h Y\@$ entail the
identity
\begin{equation*}
\hskip-2pt\Big<\!\big(d^{\,2}\Lagr\?'\big)_{\h\g}\@,\h X\otimes\h Y\Big>=N_{ij}\@X^i\@Y^j+\@G_{AB}\,U^A\@V^B\!=
\D\@\l_{\@j}/D t\@Y^j\@+\@\l_{\@j}\@\D Y^j/D t=\@\d/dt\@(\/\l_{\@j}\@Y^j\/)\@,
\end{equation*}
whence also
\begin{equation}\label{2.25}
\int_{t_0}^{\tau} \Big< \big(d^{\,2}\Lagr\?'\big)_{\@{\h\g}}\,,\,\h X\otimes\h
Y\,\Big>\,dt\ =\ Y^i\/(\tau)\,\l_{\@i}\/(\tau)\@.
\end{equation}
\end{itemize}

Let us now observe that, because of the local normality of $\g\@$, we can choose $Y\/$ in such a way as to assign to the vector $\@Y\/(\tau)\@$ whatever value
we like\@
\footnote%
{\@Notice that, in general, this may require giving up the continuity of $\@\d Y /d t$ at $\tau$.}.
In particular, introducing the notation
\begin{equation}\label{2.26}
\int_{t_0}^{\tau} \Big< \big(d^{\,2}\Lagr\?'\big)_{\@{\h\g}}\,,\,\h X\otimes\h Y\,\Big>\,dt\@:=\@a\@,\quad
\int_{t_0}^{t_1} \Big< \big(d^{\,2}\Lagr\?'\big)_{\@{\h\g}}\,,\,\h Y\otimes\h Y\,\Big>\,dt\@:=\@b\@,\quad
\end{equation}
we can always ensure the validity of the condition $\@a>0\@$.

Given any pair $X,Y$ defined as above, we now construct a $\@1$-param\-eter family of piecewise differentiable infinitesimal deformations $Z_\eta\@$,
$\@\eta\in\R\@$ according to the prescription:
\begin{equation*}
Z_\eta\/(t)\,:=\
\begin{cases}
Y\/(t)\@+\@\eta\@ X\/(t) \qquad& t_0\leq t \leq \tau\\[3pt]
Y\/(t) & \tau \leq t \leq t_1
\end{cases}
\end{equation*}

\noindent
In this way, from eqs.~\!(\ref{2.25}), (\ref{2.26}) we get the expression
\begin{multline*}
\int_{t_0}^{t_1}\!\Big<\big(d^{\,2}\Lagr\?'\big)_{\@{\h\g}}\,,\,\h Z_\eta\otimes\h Z_\eta\,\Big>\,dt\,
=\,\int_{t_0}^{\tau}\!\Big<\big(d^{\,2}\Lagr\?'\big)_{\@{\h\g}}\,,\,\eta^2\@\h X\otimes\h X\@+\@2\?\eta\@\h X\otimes\h Y\,\Big>\,dt\,+      \hskip.6cm \\
+\,\int_{t_0}^{t_1}\!\Big<\big(d^{\,2}\Lagr\?'\big)_{\@{\h\g}}\,,\,\h Y\otimes\h Y\,\Big>\,dt\,=\,2\?a\?\eta\,+\,b\hskip.3cm
\end{multline*}
which is \emph{negative\/} for $\@\eta<-\@b/2\?a\@$.

On the other hand, as proved in Appendix \ref{SA}, given any piecewise differentiable $\@Z_\eta\@$ there exists a differentiable infinitesimal deformation
$Z_\eta^{\@\prime}$ such that the difference
\begin{equation*}
\int_{t_0}^{t_1} \Big< \big(d^{\,2}\Lagr\?'\big)_{\@{\h\g}}\,,\@\h Z_\eta^{\@\prime}\otimes\h Z_\eta^{\@\prime}\,\Big>\,dt -
\int_{t_0}^{t_1} \Big< \big(d^{\,2}\Lagr\?'\big)_{\@{\h\g}}\,,\@\h Z_\eta\otimes\h Z_\eta\,\Big>\,dt
\end{equation*}
is as small as we wish.

Collecting all results we conclude that, whenever a point $\tau\in(\@t_0,t_1\?)\@$ conjugate to $t_0\@$ exists, there is at least one differentiable
infinitesimal deformation $Z\@'$ vanishing at the endpoints of $\g$ and satisfying $\@\int_{t_0}^{t_1}\big<\big(d^{\,2}\Lagr\?'\big)_{\@{\h\g}}\@,\@\h
Z^{\@\prime}\otimes\h Z^{\@\prime}\@\big>\,dt\@<\@0\@$.

This proves assertion (i). Assertion (ii) does not require any additional proof, but is merely a restatement of a result established in Theorem \ref{Teo2.1}.
\end{proof}
\section{Jacobi fields}\label{S3}
A deeper insight into the concept of conjugate points comes from the study of the \emph{Jacobi vector fields\/}.
The idea is well known: given a regular, locally normal extremal $\@\g\colon[t_0,t_1]\to\V\@$ of the action functional $\@\I\@[\g]\@$, we focus on a special
class of deformations $\@\g_{\@\xi}\@$ consisting of $\@1\@$--parameter families of \emph{extremals\/} of $\@\I\@[\g]\@$. No restriction is posed on the
behaviour of the endpoints $\@\g_{\@\xi}\/(t_0)\@,\,\g_{\@\xi}\/(t_1)\@$.

Given any such $\@\g_{\@\xi}\@$, we preserve the notation $\@X=X^i\@\big(\de/de{q^i}\big)_\g\@$ for the associated infinitesimal deformation,
$\h\g\colon[t_0,t_1]\to\A\@$ for the lift of $\g\@$ to a section of the velocity space and\vspace{.6pt}\linebreak
 $\@\h X=X^i\@\big(\de/de{q^i}\big)_{\h\g}\@
+\@X^A\@\big(\de/de{z^A}\big)_{\h\g}\@$ for the lift of $\@X\@$ to a vector field along $\h\g\@$.

The whole setup is transferred to the environment $\C\/(\A)$, denoting by $\gt$ the (unique) extremal of the functional
$\@\int_{\gt}\@\th_{\hskip-.5pt\Lagr}\@$\vspace{1pt} projecting onto $\g$ and considering deformations $\@\gt_{\@\xi}\@$ consisting of $1$--parameter families
of extremals of $\@\int_{\gt}\@\th_{\hskip-.5pt\Lagr}\@$\vspace{1pt}.

As pointed out in Sec.\!~\ref{S2}, the procedure is not gauge--invariant, but explicitly depends on the choice of $\@\Lagr$.
In particular, replacing the original Lagrangian with an adapted one $\@\Lagr\?'=\Lagr-\dot S$ yields a setup that, without affecting the essence of the
problem, ensures the vanishing of the functions $\@p_i\/(t)\@$ along $\@\gt\@$, thus entailing the identification $\@\gt=\O\cdot\h\g\@$.

In coordinates, sticking to the stated choice of $\@\Lagr\?'$ and adopting the representation
\begin{equation*}
\gt_{\?\xi}:\qquad q^i=\@\vphi^i\/(\xi,t)\,,\quad\; z^A=\@\z^A\/(\xi, t)\,,\quad\;p_i\,=\@\rho_i\/(\xi,t)\,,
\end{equation*}
the request that each section $\@\gt_{\?\xi}:[\?t_0,t_1\@]\to\C\/(\A)\@$ be an extremal of the functional $\@\int_{\gt}\@\th_{\hskip-.5pt\Lagr\?'}\@$ is
summarized into the Pontryagin equations\@
\footnote%
{\@As a check of inner consistency it may be noticed that, in view of the condition $(d\/\Lagr\?')_{\h\g}=0\@$, eqs.~\!(\ref{3.1}\@b,\@c) and the normality of
$\@\g\@$ yield back the relation $\@\rho_i\/(0,t)=0\@$.}

\begin{subequations}\label{3.1}
\begin{align}
 & \de\vphi^i/de t\,=\,\psi^i(t,\vphi^i,\z^A)\,,                                            \\[2pt]
 & \de \rho_i/de t+\de\psi^k/de{q^i}\,\rho_k\,=\,\de \Lagr\?'/de{q^i}\,,                    \\[3pt]
 & \rho_i\,\de\psi^i/de{z^A}\,=\, \de \Lagr\?'/de{z^A}\,.\vspace{4pt}
\end{align}
\end{subequations}

\medskip \noindent
Let $\@\tilde X= X^i\big(\de/de{q^i}\big)_{\gt}+\@X^A\big(\de/de{z^A}\big)_{\gt} +\@\l_{\@i}\big(\de/de{p_i}\big)_{\gt}\@$\vspace{1pt} denote the infinitesimal
deformation associated to $\@\gt_{\@\xi}\@$, with
\begin{equation}\label{3.2}
X^i=\,\bigg(\de\?\vphi^i/de\xi\bigg)_{\!\xi=0}\;,\quad X^A=\,\bigg(\de\@\z^A/de\xi\bigg)_{\!\xi=0}\;,
\quad\l_{\@i}\,=\,\bigg(\de\?\rho_i/de\xi\bigg)_{\!\xi=0}\;.
\end{equation}
Taking eqs.~\!(\ref{3.1}) and the relation $\@\rho_i\/(0,t)=0\@$ into account, it is easily seen that the components (\ref{3.2}) satisfy the following system
of differential--algebraic equations
\begin{subequations}\label{3.3}
\begin{align}
&\d X^i/d t\,=\,\bigg(\de\psi^i/de{q^k}\bigg)_{\!\h\g}\@X^k\@+\@\bigg(\de\psi^i/de{z^A}\bigg)_{\!\h\g}\@X^A,                            \\[5pt]
&\d\l_{\@i}/dt\,+\,\l_{\@k}\bigg(\de\psi^k/de{q^i}\bigg)_{\!\h\g}\@=\,
\bigg(\sd\Lagr\?'/de q^i/de{q^k}\bigg)_{\!\h\g}\@X^k \@+\@\bigg(\sd\Lagr\?'/de q^i/de{z^A}\bigg)_{\!\h\g}\@X^A,                        \\[5pt]
&\l_{\@i}\bigg(\de\@\psi^{\?i}/de{z^A}\bigg)_{\!\h\g}\,=\,\bigg(\sd\Lagr\?'/de z^A/de{q^k}\bigg)_{\!\h\g}\@X^k\@+\@
\bigg(\sd\Lagr\?'/de z^A/de{z^B}\bigg)_{\!\h\g}\@X^B.
\end{align}
\end{subequations}

Eqs.~(\ref{3.3}\@a,\@b,\@c) nearly resemble the Pontryagin ones (\ref{1.20}). To pursue this viewpoint, we regard $\@\Vg\@$ as the configuration manifold of an
abstract system $\@\B\?'$ and $\@\Ag\to\Vg\@$ as the associated space of admissible velocities, thus establishing a bijective correspondence between the
admissible evolutions of $\@\B\?'\@$ and the infinitesimal deformations of $\@\g\@$.

Introducing coordinates $\@t,u^i\@$ in $\@\Vg\@$ and $\@t,u^i,v^A\@$ in $\@\Ag\@$ according to the prescriptions (\ref{1.5}\@a,\@b), the imbedding
$\@i_*\colon\Ag\to\j\Vg\@$ is locally expressed by eq.~\!(\ref{1.6}b), now synthetically written as
\begin{equation*}
\dot u^i=\left(\de\psi^i/de{q^k}\right)_{\!\h\g}\@u^k + \left(\de\psi^i/de{z^A}\right)_{\!\h\g}\@v^A\,:=\,\Psi^i\/(t,u^i,v^A)\@.
\end{equation*}

\noindent
The picture is completed adopting the quadratic form
\begin{equation}\label{3.4}
\mathfrak{L}\/(\h X)\,:=\,\frac12\,\Big<\big(d^{\,2}\Lagr\?'\big)_{\@\h\g}\,,\,\h X\otimes\h X\@\Big>
\end{equation}
as a Lagrangian on $\@\Ag\@$ and denoting by $\@\mathfrak{I}\@$ the functional assigning to each admissible section $\@X\colon[\@t_0,t_1]\to\Vg\@$ the action
integral $\@\mathfrak{I}\,[\?X\?]:=\int_{\h X}\mathfrak{L}\,\@d\/t\@$.
In this way, for each finite deformation $\@\g_{\?\xi}\@$ of $\@\g\@$ tangent to $\@X\@$,\vspace{1pt} eqs.~\!(\ref{2.16}), (\ref{3.4}) provide the
identification\linebreak $\@\mathfrak{I}\,[\?X\?]=
\frac12\,\frac{d^{\@2}\I\?[\g_\xi]}{d\?\xi^{\?2}}\@\big|_{\?\xi=0}\@$.

\noindent
In coordinates, eq.~\!(\ref{3.4}) reads
\begin{equation*}
\mathfrak{L}\/(t,u^i,v^A)=\frac12\bigg[\bigg(\sd\Lagr\?'/de q^i/de{q^j}\bigg)_{\!\h\g}u^i\@u^j
+2\@\bigg(\sd\Lagr\?'/de q^i/de{z^A}\bigg)_{\!\h\g}u^i\@v^A+
\bigg(\sd\Lagr\?'/de z^A/de{z^B}\bigg)_{\!\h\g}v^A\@v^B\,\bigg]\@.
\end{equation*}
The Pontryagin equations for the determination of the extremals of the functional $\@\mathfrak{I}\,[\?X\?]\@$ take therefore the form
\begin{subequations}\label{3.5}
\begin{align}
& \d X^i/d t\,=\,X^k\, \de\Psi^i/de{u^k} \, +\, X^A\, \de\Psi^i/de{v^A}                                \\[5pt]
&\d\l_{\@i}/dt\,+\,\l_{\@k}\,\de\Psi^k/de{u^i}\@=\,\de {\mathfrak{L}} /de {u^i}                          \\[5pt]
&\l_{\@i}\;\de\Psi^i/de{v^A}\,=\,\de {\mathfrak{L}} /de {v^A}\@.
\end{align}
\end{subequations}
rephrasing in different notation the content of eqs.~\!(\ref{3.3}\@a,\@b,\@c).

\begin{definition}\label{Def3.1}
The variational problem based on the Lagrangian (\ref{3.4}) is called the \emph{accessory variational problem\/}. The extremals of the functional
$\@\mathfrak{I}\,[\?X\?]\@$ are called the Jacobi vector fields along $\@\g\@$.
\end{definition}

\noindent
Collecting all results we conclude
\begin{proposition}\label{Pro3.1}
Every infinitesimal deformation arising from a finite deformation $\@\g_\xi\@$ consisting of a $\@1\@$--parameter family of extremals is a Jacobi vector field
along $\@\g\@$.
\end{proposition}

Notice that the previous argument do not ensure that \emph{every\/} Jacobi vector field is related to a corresponding $\@1\@$--parameter family of extremals
$\@\g_\xi\@$ in the way described in Proposition \ref{Pro3.1}. However, this is not a crucial issue: what really matters is establishing a relationship between
the solutions of the accessory variational problem and the second variation $\@\frac{d^{\@2}\I\?[\g_\xi]}{d\?\xi^{\?2}}\@\big|_{\?\xi=0}\@$.\vspace{1pt}

To this end we stick to the original formulation, and regard eqs.~\!(\ref{3.3}\@a,\@b,\@c) as a system of differential--algebraic equations for the
determination of a vector field $\@\tilde X\@$ along $\@\gt\@$. Recalling Remark \ref{Rem1.1} we next observe that, due to the identification
$\@\gt=\O\cdot\h\g\@$, the unknown $\@\tilde X\@$ may be resolved into a pair
\begin{equation*}
\h X\,=\,X^i\@\bigg(\de/de{q^i}\bigg)_{\h\g}+\,X^A\@\bigg(\de/de{z^A}\bigg)_{\h\g}\,,\qquad\;
\h\l\,=\,\l_{\@i}\;\big(\@d\/q^i-\psi^i\@d\/t\@\big)_{|\?\h\g}\,,
\end{equation*}
consisting of a vector field and a contact $1$--form along the projected curve $\@\h\g=\zeta\,\cdot\gt\@$.

In turn, $\@\h X\@$ is the lift of a vector field $\@X:=X^i\@\big(\de/de{q^i}\big)_{\g}\@$, clearly recognized as the infinitesimal deformation associated with
the $\@1\@$--parameter family of sections $\@\g_\xi=\pi\,\cdot\zeta\,\cdot\gt_{\@\xi}\@$, while $\@\h\l\@$ determines (and is completely determined by) a
virtual $1$--form $\@\l=\l_{\@i}\;\wg{i}\@$ along $\@\g\@$.

Finally, under the (\@crucial\@) hypothesis of regularity of $\@\g\@$, we can use the infinitesimal control $\@h\colon\Vg\to\Ag\@$ induced by the Lagrangian
$\@\Lagr\?'\@$ to split the field $\@\hat{X}\@$ into a horizontal and a vertical part according to the prescription
\begin{equation*}
\h X\,=\,h\/(X)\@+\@U\,=\,X^i\,\DE_i\@+\@U^A\@\biggl(\de/de{z^A}\biggr)_{\h\g}
\end{equation*}
with
\begin{equation*}
U^A\@=\,X^A\@+\,G^{AB}\,\biggl(\sd\Lagr\?'/de q^i/de{z^B}\biggr)_{\!\h\g}\@X^i\@.
\end{equation*}

\noindent
On account of the identification $\@G_{AB}=\big(\sd\Lagr\?'/de z^A/de{z^B}\big)_{\h\g}\@$, this allows to cast eq.~\!(\ref{3.3}c) into the form
\begin{equation*}
\l_{\@i}\bigg(\de\@\psi^{\?i}/de{z^A}\bigg)_{\!\h\g}\,=\,G_{AB}\;U^B \quad\Longrightarrow\quad
U^A\,=\,G^{AB}\,\l_{\@i}\,\biggl(\de\@\psi^{\?i}/de{z^B}\biggr)_{\!\h\g}\,,
\end{equation*}
mathematically equivalent to the linear relation
\begin{equation}\label{3.6}
X^A\@=\,G^{AB}\@\biggl[\@\biggl(\de\@\psi^{\?i}/de{z^B}\biggr)_{\!\h\g}\@\l_{\@i}\,-\,\biggl(\sd\Lagr\?'/de q^i/de{z^B}\biggr)_{\!\h\g}\@X^i\@\biggr]\@.
\end{equation}

Substituting eq.~\!(\ref{3.6}) into eqs.~\!(\ref{3.3}\@a,\@b), recalling the definitions of the tensors $\@N_{ij}\@$, $M^{ij}$ and expressing the ordinary time
derivatives in terms of the absolute ones, we eventually obtain the system of differential equations
\begin{subequations}\label{3.7}
\begin{align}
&\D X^i/Dt\,=\,G^{AB}\,\bigg(\de\psi^i/de{z^A}\bigg)_{\!\h\g}\bigg(\de\psi\?^j/de{z^B}\bigg)_{\!\h\g}\,\l_{\@j}\,=\,M^{\@ij}\,\l_{\@j}          \\[4pt]
& \D\@\l_{\@i}/Dt\,=\bigg[\bigg(\sd\Lagr\?'/de q^i/de{q^j}\bigg)_{\!\h\g}\!-\,
G^{AB}\bigg(\sd\Lagr\?'/de q^i/de{z^A}\bigg)_{\!\h\g}\bigg(\sd\Lagr\?'/de q^j/de{z^B}\bigg)_{\!\h\g}\,\bigg]\,X^j=\,N_{ij}\,X^j
\end{align}
\end{subequations}
formally identical to the system (\ref{2.22}\@a,\@b) encountered in the proof of Theorem \ref{Teo2.1}.

An alternative derivation of eqs.~\!(\ref{3.7}) is obtained following a procedure analogous to the one outlined in Sec.~\!\ref{S1.6} in order to cast the
Pontryagin equations (\ref{3.5}) into Hamiltonian form.
In this way, eqs.~\!(\ref{3.7}) are recognized as the Hamilton equations associated with the Hamiltonian
\begin{equation}\label{3.8}
\mathfrak{H}\,=\,\frac12\,M^{ij}\@\l_{\@i}\@\l_{\@j}\,-\,\frac12\,N_{ij}\@X^i\@X^j\@-\,\tau\?_k{}^i\,X^k\@\l_{\@i}\@,
\end{equation}
$\@\tau\?_k{}^i\@$ being the temporal connection coefficients involved in the definition of the absolute time derivative $\@\D/Dt\@$.
As a check, we let the reader verify that the right-hand side of (\ref{3.8}) is indeed identical to the difference $\@\l_{\@i}\@\Psi^i-\@\mathfrak L\,$,
restricted to the $\@2n$--dimensional subbundle described by eq.~\!(\ref{3.5}\@c).

\begin{remark}\label{Rem3.1}
As usual, the algorithm gets simplified referring all tensors to an $h$--transported basis $\@\big\{\@e^\na,\@e_\na\@\big\}\@$. In this way, eqs.~\!(\ref{3.7})
take the form
\begin{equation*}
\d X^\na/dt\,=\,M^{\na\nb}\,\l_{\@\nb}\,,\qquad \d\l_{\@\na}/dt\,=\,N_{\na\nb}\,X^\nb,
\end{equation*}
while the Hamiltonian simplifies to
\begin{equation*}
\mathfrak{H}\@'\,=\,\frac12\,M^{\na\nb}\@\l_{\@\na}\@\l_{\@\nb}\,-\,\frac12\,N_{\na\nb}\@X^\na\@X^\nb\@
\end{equation*}
Once again, we let the reader verify that the relation $\@\mathfrak{H}\@'=\mathfrak{H}\@+\@\tau\?_k{}^i\,X^k\@\l_{\@i}\@$ reflects the transformation law of
the Hamiltonian under arbitrary changes of the independent coordinates.
\end{remark}

\noindent
Collecting all results and recalling Definition \ref{Def2.1} we can eventually state
\begin{proposition}\label{Pro3.2}
A point $\@\g\/(\tau)\@$, $\@\tau>t_0\@$ along a regular, locally normal extremal\linebreak $\@\g\colon[\@t_0,t_1]\to\V\@$ is conjugate to $\@\g\/(t_0)\@$ if and only if
there exists a non--zero Jacobi vector field $\@X\colon[\@t_0,t_1]\to\Vg\@$ satisfying $\@X\/(t_0)=X\/(\tau)=0\@$.
\end{proposition}
\medskip
\appendix
\section{A smoothing theorem}\label{SA}
Let $\@\g\colon[\@t_0,t_1]\to\V\@$ be an admissible section carrying a non--singular matrix $G_{AB}$. Given a piecewise differentiable infinitesimal
deformation $\@X\@$ of $\@\g\@$ vanishing at the endpoints, we want to build a $1$--parameter family of differentiable infinitesimal deformations $X_{\eta}\@$
vanishing at the endpoints and satisfying
\begin{equation}\label{A.1}
\lim_{\eta\to0}\ \int_{\h\g} \Big< \big(d^{\,2}\Lagr\?'\big)_{\@{\h\g}}\,,\,\h {X}_\eta\otimes\h{X}_\eta\,\Big>\,dt\,=\,
\int_{\h\g} \Big< \big(d^{\,2}\Lagr\?'\big)_{\@{\h\g}}\,,\,\h {X}\otimes\h {X}\,\Big>\,dt\@.
\end{equation}
For the present purposes, we shall concentrate on a single discontinuity, located at $t=t^*$.
To start with, we fix the gauge in such a way as to ensure that the expression
\begin{equation}\label{A.2}
\Big<\big(d^{\,2}\Lagr\?'\big)_{\@{\h\g}}\,,\,\h Y\otimes\h Z\,\Big>\,=\,G_{AB}\ Y^A\, Z^B
\end{equation}
holds for all $\@t\@$ in some closed interval $\@[\?t^*,t^*\!+\eps\?]\@$, $\@(\eps>0)\@$ and all $\@\h Y,\h Z\in\A\/(\h\g\/(t))\@$.

We next stick to the algorithm illustrated in Sec.\;\ref{S1.5}, identifying $\@h\@$ with the infinitesimal control (\ref{2.9}) induced by $\@\Lagr\?'$. On
account of eq.~\!(\ref{1.14}), every infinitesimal deformation\linebreak $\@X=X^i\@\big(\de/de{q^i}\big)_\g\@$\vspace{1pt} vanishing at $\@t_0\@$ is then determined by a
corresponding vertical vector field $\@U=U^A\big(\de/de{z^A}\big)_{\h\g}\@$ through the relation
\begin{equation}\label{A.3}
X^i\/(t)\,=\biggl(\@\int_{t_0}^{t}\@\psi^\na_{\,A}\,U^A\,d\/\tau\biggr)\,e_\na^{\;\,i}\/(t)\@.
\end{equation}
In particular:
\begin{itemize}
\item
possible discontinuities in the first derivatives $\@\d X^i/dt\@$ at $\@t=t^*\@$ are reflected into corresponding discontinuities of the components
$U^A\/(t)\@$. As usual, these will be dealt with prolonging both restrictions $\@U\raise-2pt\hbox{$|$}{\plus02}_{[t_0,t^*)}\@$,
$\@U\raise-2pt\hbox{$|$}{\plus02}_{[t^*,t_1,)}\@$ to differentiable vector fields $\@U_\narc{-}\,$ and $U_\narc{+}\@$ along $\h\g\@$;
\item
the request for the vanishing of $X$ at $\@t=t_1\@$ results in the condition
\begin{equation}\label{A.4}
\int_{t_0}^{t_1} \psi^\na_{\,A}\ U^A\, d\/t\ =\ 0\,;
\end{equation}
\item \vspace{2pt}
the lift of $\@X\@$ is given by $\@\h X=h\/(X)+U=X^i\@\DE_i\@+\@U^A\big(\de/de{z^A}\big)_{\h\g}\@$.
\end{itemize}

\noindent
After these preliminaries, let us now introduce $\@n\@$ differentiable functions along $\@\g\@$ according to the prescription
\begin{equation}\label{A.5}
f^\ni\/(t)\@:=\@\psi^\ni_{\,A}\/(t)\/\Big[\@U^A_\narc{-}\/(t)-U^A_\narc{+}\/(t)\@\Big]\@,\qquad t_0\le t\le t_1
\end{equation}
and denote by $\vphi^\a,\,\a=1\And k\@$ a (possibly smaller) subfamily such that the restrictions $\@\vphi^\a\/(t)\@$ to the interval $\@[\?t^*,t^*+\eps\?]\@$
form a basis for the linear space spanned by the restrictions of the functions (\ref{A.5}) to the same interval.
Setting $\tau:=t-t^*$, we have then the representation
\begin{equation*}
f^\ni\/(\tau)\,=\,\sum_{\a=1}^k\mathscr{A}^\ni_{\;\,\a}\,\vphi^\a\/(\tau)\@,\qquad\;0\le\tau\le\eps\,,
\end{equation*}
with $\@\rank\@\mathscr{A}^\ni_{\;\,\a}=k$.

We next consider $k\@$ further differentiable functions $g_\b\/(\tau)\@$, with support contained in $(\@0\@,\eps)$, satisfying the requirement that the matrix
\begin{equation}\label{A.6}
 \mathscr{B}^{\@\a}_{\;\,\b}\,:=\,\int_0^\eps\vphi^\a\/(\tau)\ g_\b\/(\tau)\,d\/\tau
\end{equation}
be non--singular. With the stated notations, this entails the relation
\begin{equation*}
\int_0^\eps f^\ni\/(\tau)\ g_\b\/(\tau)\, d\/\tau\ =\ \sum_ {\a=1}^k \mathscr{A}^\ni_{\;\,\a}\, \mathscr{B}^{\@\a}_{\;\,\b}\ =\
\big(\@\mathscr{A}\/\cdot\/\mathscr{B}\@\big)^\ni_{\;\,\b}\,.
\end{equation*}
Finally, let $g_0\/(\vsigma)$ be a differentiable function over $\@\R\@$ fulfilling the conditions
\begin{equation*}
\left\{
\begin{alignedat}{2}
 &g_0\/(\vsigma)\,=\,1 &&\forall\;\vsigma\leq\@ 0                               \\
 & 0\le g_0\/(\vsigma)\le 1 \qquad &&\forall\;\vsigma\in[\@0,1\@]               \\
 &g_0\/(\vsigma)\,=\,0\qquad&&\forall\;\vsigma\geq\@ 1
\end{alignedat}
\right.
\end{equation*}

\noindent
For any $\@\eta\in(0,\?\eps\?)\@$, $\@\underline{\xi}=(\xi^1\And\xi^k)\in\R^k\@$, the expression
\begin{equation}\label{A.7}
\tilde{g}\@(\tau,\eta,\underline{\xi}):=\ g_0\/\bigg(\frac\tau\eta\bigg)\,-\,\eta\,\sum_{\b=1}^k\,\xi^\b\@g_\b\/(\tau)
\end{equation}
is then a differentiable function of its arguments, satisfying $\@\tilde{g}\@(\tau,\eta,\underline{\xi})=1\@$ $\,\forall\,\tau<0\@$ and
$\,\tilde{g}\@(\tau,\eta,\underline{\xi})=0\@$ $\,\forall\,\tau>\eps\@$. Bearing all this in mind, we now state
\begin{thm}
In the parameters space, there exists a unique curve $\underline{\xi}=\underline{\xi}\/(\eta)\@$ such that, setting
\begin{equation}\label{A.8}
U^A_\eta\/(t)\/:=\@\tilde{g}\?\big(t-t^*\!,\@\eta\@,\@\underline{\xi}\/(\eta)\/\big)\@\big[\@U^A_\narc{-}\/(t)\@-U^A_\narc{+}\/(t)\@\big]\@+\@ U^A_\narc{+}\/(t)\@,
\end{equation}
the fields $U^A_\eta\big(\de/de{z^A}\big)_{\h\g}\@$ generate a $1$--\@parameter family of differentiable infinitesimal deformations $X_\eta$ vanishing at the
endpoints of $\@\g\@$ and fulfilling the condition \eqref{A.1}.
\end{thm}
\begin{proof}
For each differentiable curve $\underline{\xi}=\underline{\xi}\/(\eta)\@$, the functions (\ref{A.8}) are differentiable and satisfy
$\@U^A_\eta\/(t)=U^A_\narc{-}\/(t)\@$ for $\@t\le t^*\@$ and $\@U^A_\eta\/(t)=U^A_\narc{+}\/(t)\@$ for $\@t\ge t^*+\eps\@$.

A necessary and sufficient condition for them to determine an infinitesimal deformation vanishing at the endpoints is that they fulfil the requirement
(\ref{A.4}).
Since the original functions $U^A\/(t)\@$ already do, this means requiring the validity of the relation
\begin{equation*}
\int_{t_0}^{t_1}\!\psi^\na_{\,A}\,\big(\@U^A_\eta\?-\?U^A\big)\@d\/t\@=\!
\int_{0}^{\eps}\!\psi^\na_{\,A}\,\tilde{g}\/\big(\tau,\@\eta\@,\@\underline{\xi}\/(\eta)\/\big)
\Big[\@U^A_\narc{-}\/(t^*\!+\tau)-U^A_\narc{+}\/(t^*\!+\tau)\@\Big]\@d\/\tau=0\?.
\end{equation*}
On account of eqs.~\!(\ref{A.5}), (\ref{A.6}), the latter may be written as
\begin{equation*}
\frac1\eta\,\int_0^\eps g_0\/\bigg(\frac\tau\eta\bigg)\,\vphi^\a\/(\tau)\,d\/\tau\,=\,\sum_{\b=1}^k\int_0^\eps\xi^\b\,\vphi^\a\/(\tau)\,g_\b\/(\tau)\,d\/\tau
\end{equation*}
or also, denoting by $\mathscr{C}^\a\/(\eta)\@$ the left--hand term and recalling eq.~\!\eqref{A.6},
\begin{equation}\label{A.9}
\mathscr{C}^{\@\a}\/(\eta)\,=\,\sum_{\b=1}^k\,\mathscr{B}^{\@\a}_{\;\,\b}\,\@\xi^\b.
\end{equation}

Due to the non--singularity of the matrix $\mathscr{B}^{\@\a}_{\;\,\b}\@$, eq.~\!(\ref{A.9}) uniquely determines the coefficients $\xi^1,\ldots,\xi^k\@$ in
terms of $\@\mathscr{C}^{\@\a}\/(\eta)\@$. In particular, the conditions $\@\eta<\eps\@$, $g_0\/(\vsigma)=0$ $\@\forall\,\vsigma\geq1\@$ entail the
identifications
\[
\frac1\eta\,\int_0^\eps g_0\/\bigg(\frac\tau\eta\bigg)\,\vphi^\a\/(\tau)\,d\/\tau\,=\,\frac1\eta\,
\int_0^\eta g_0\/\bigg(\frac\tau\eta\bigg)\,\vphi^\a\/(\tau)\,d\/\tau\,=\,\int_0^1 g_0\/(\vsigma)\,\@\vphi^\a\/(\eta\@\@\vsigma)\,d\/\vsigma\@,
\]
indicating that the functions $\@\mathscr{C}^{\@\a}\/(\eta)\@$ converge to a finite limit when $\eta\to 0\@$.

Summing up we conclude that, for any $\eta\in(\@0\@,\eps)\@$, the request that $U^A_\eta\/(t)\big(\de/de{z^A}\big)_{\h\g}\@$ generates an infinitesimal
deformation $\@X_\eta\@$ vanishing at the endpoints uniquely determines the functions $\xi^\a\/(\eta)$, ensuring as well their \emph{boundeness\/} in the limit
$\eta\to 0\@$.

\smallskip
Let us finally establish eq.~\!(\ref{A.1}). To this end, we lift both infinitesimal deformations $\@X_\eta\@$,~$X\@$ to~corresponding deformations $\h
X_\eta\@,\,\h X\@$ of the section $\@\h\g\@$ and notice that, on account of eqs.~\!(\ref{A.3}), (\ref{A.8}), the difference $\h X_\eta-\h X\@$ vanishes outside
the interval $\@(\@t^*\!,\@t^*\!+\@\eps\@)\@$.
Together with eq.~\!(\ref{A.2}), reflecting the stated choice of $\@\Lagr\?'$, this entails the evaluation\vspace{4pt}
\begin{equation}\label{A.10}
\begin{split}
\int_{\h\g} \Big<& \big(d^{\,2}\Lagr\?'\big)_{\@{\h\g}}\,,\,\h {X}_\eta\otimes\h
{X}_\eta\,\Big>\,d\/t - \int_{\h\g} \Big< \big(d^{\,2}\Lagr\?'\big)_{\@{\h\g}}\,,\,\h {X}\otimes\h {X}\,\Big>\,d\/t =                               \\[0pt]
&= \int_{0}^\eps \!\!\Big< \big(d^{\,2}\Lagr\?'\big)_{\@{\h\g}}\,,\,\big(\/\h {X}_\eta + \h X\/\big)\otimes
\big(\/\h{X}_\eta - \h X\/\big)\,\Big>\,d\/\tau =                                                                                                   \\[4pt]
&= \int_0^\eps \!\!G_{AB}\@ \big(\/U^A_\eta + U^A_\narc{+}\?\big)\@\big(\/ U^A_\eta - U^A_\narc{+}\?\big)\,d\/\tau =                                \\[4pt]
&= \int_0^\eps \!\!G_{AB}\@ \big(\/U^A_\eta+U^A_\narc{+}\?\big)\@
\big(\/ U^A_\narc{+} - U^A_\narc{-}\?\big)\,\tilde{g}\/\big(\@\tau,\eta,\underline{\xi}\/(\eta)\@\big)\,d\/\tau\@.
\end{split}
\end{equation}

\medskip
Observing that the expression $\nu\/(\tau,\eta):=G_{AB}\@ \big(\/U^A_\eta +U^A_\narc{+}\/\big)\@\big(\/ U^A_\narc{+}-U^A_\narc{-}\/\big)\@$ is bounded for
$\eta\to0$ and taking eq.~(\ref{A.7}) into account, we conclude the last--hand term in eq.~\!(\ref{A.10}), suitably rewritten as
\begin{equation*}
\begin{split}
\int_0^\eps \nu\/(\tau,\eta)\@ &\bigg[\@g_0\/\bigg(\frac\tau\eta\bigg)-\@\eta\,\sum_{\b=1}^k\,\xi^\b\/(\eta)\,\@g_\b\/(\tau)\@\bigg]\,d\/\tau\,=        \\
=\eta&\bigg[\@\int_0^1\nu\/(\vsigma\@\eta,\eta)\,g_0\/(\vsigma)\,d\/\vsigma\,-\,
\sum_{\b=1}^k\,\xi^\b\/(\eta)\@\int_0^\eps \nu\/(\tau,\eta)\,\@g_\b\/(\tau)\,d\/\tau\@\bigg]
\end{split}
\end{equation*}
is infinitesimal in the limit $\eta\to0\@$.
\end{proof}


\bigskip
\begin{thebibliography}{99}
\bibitem{mbp1} E.~Massa, D.~Bruno and E.~Pagani, Geometric control theory I: mathematical foundations,
               \emph{arXiv}:0705.2362v2 [math.OC]

\bibitem{BLP} D.~Bruno, G.~Luria and E.~Pagani, On the gauge structure of the calculus of variations with constraints, 
              \emph{Int. J. Geom. Methods Mod. Phys.} {\bf 08}, 1723--1746 (2011).

\bibitem{Sternberg} S. Sternberg, {\it Lectures on Differential Geometry\/},
               Prentice Hall, Englewood Cliffs, New Jersey (1964).  
               
               
\bibitem{Warner} F.~W.~Warner, {\it Foundations of Differential Manifolds and Lie Groups\/},
         Springer--Verlag, New York (1983).


\bibitem{MP2} E.~Massa and E.~Pagani, A new look at Classical Mechanics of
              constrained systems,
              {\it Ann.~Inst.~Henri Poincar\'e}, Physique th\'eorique,
              {\bf 66}, 1--36, (1997).
                              
\bibitem{Saunders} D.J.~Saunders, {\it The Geometry of Jet Bundles\/},
              London Mathematical Society, Lecture Note Series 142,
              Cambridge University Press (1989).
              
\bibitem{Pom} J.F.~Pommaret, {\it Systems of Partial Differential Equations and Lie
               Pseudogroups}, Gordon \& Breach, New York (1978).
       

\bibitem{DeLeon} M.~de Leon and P.R.~Rodrigues, {\it Methods of Differential Geometry
              in Analytical Mechanics\/}, North Holland, Amsterdam (1989).

\bibitem{Bliss} G.A.~Bliss, {\it Lectures on the calculus of the variations\/},
              The University of Chicago Press, Chicago (1946).
              
\bibitem{Lanczos} C.~Lanczos, {\it The variational principles of mechanics\/},
               University of Toronto Press, Toronto (1949)
               (Reprinted by Dover Publ. (1970)).              
              
\bibitem{Gelfand} I.M.~Gelfand  and S.V.~Fomin, {\it  Calculus of variations\/},
               Prentice-Hall Inc., Englewood Cliffs (1963).
                       
\bibitem{Giaquinta} M.~Giaquinta and S.~Hildebrandt, {\it Calculus of variations I, II\/},
               Springer-Verlag, Berlin Heidelberg New York (1996).

\bibitem{Rund} H.~Rund, {\it The Hamilton-Jacobi theory in the calculus of variations\/},
               Van Nostrand, London (1966).

\bibitem{Hestenes} M.R.~Hestenes, {\it Calculus of variations and optimal control theory\/},
               Wiley, New York London Sydney (1966).
               
\bibitem{Sagan} H.~Sagan, {\it Introduction to the calculus of variations}, McGraw--Hill Book Company, New York (1969)               
               
\bibitem{Pontryagin} L.S.~Pontryagin, V.G.~Boltyanskii, R.V.~Gamkrelidze and E.F.~Mishchenko,
               {\it The mathematical theory of optimal process\/},
               Interscience, New York (1962).
              
\bibitem{Young} L.~C.~Young {\it Lectures on the Calculus of Variations and Optimal Control
               Theory\/} (second edition), AMS Chelsea Publishing, New York (1980).

\bibitem{Griffiths} P.~Griffiths, {\it Exterior differential systems and the calculus of
               variations\/}, Birkhauser, Boston (1983).
               
\bibitem{Sussmann}  H.J.~Sussmann, An introduction to the coordinate-free maximum principle, 
                 in {\it Geometry of Feedback and Optimal Control}, 
                 Eds. B.~Jakubczyk and W.~Respondek, 
                 Marcel Dekker, New York, 463--557 (1997).  
                 
\bibitem{Respondek}  W.~Respondek, Introduction to geometric nonlinear control; linearization, 
                 observability and decoupling, in {\it Mathematical Control Theory} (Ed. A.~Agrachev), 
                 ICTP Lecture Notes, 169--222,(2002).            
               
\bibitem{Montgomery} R.~Montgomery, {\it A Tour of Subriemannian Geometries, Their
           Geodesics and Applications}, AMS, Math. Surveys and Monographs, Vol. 91 (2000).

\bibitem{Agrachev1} A.~A.~Agrachev and Yu.L.~Sachov, {\it Control Theory from the Geometric
               Viewpoint\/}, Springer-Verlag, Berlin Heidelberg New York (2004).

\bibitem{Radon1} J.~Radon, \"Uber die Oszillationstheoreme der konjugierten Punkte biem 
                Probleme von Lagrange,
                {\it M\"unchner Sitzungsberichte}, 243--257 (1927).

\bibitem{Radon2} J.~Radon, Zum Problem von Lagrange,
                 {\it Abhandlungen aus dem Mathematischen Seminar Hamburg} {\bf 6}, 273--299 (1928).  

\bibitem{Reid1} W.H.~Reid, A Matrix Differential Equation of Riccati Type,
                {\it Amer. J. Math.} {\bf 68}, 237--246 (1946).  
\end{thebibliography}
\end{document}